	\documentclass[12pt,preprint]{aastex}
	\shorttitle{Be Star Disks}
	\shortauthors{Gies et al.}
\begin{document}

\received{2006 June 21}

\title{CHARA Array $K^\prime$-band Measurements of the \\
Angular Dimensions of Be Star Disks}

\author{D. R. Gies\altaffilmark{1}, 
 W. G. Bagnuolo, Jr.,
 E. K. Baines, 
 T. A. ten Brummelaar, 
 C. D. Farrington, 
 P. J. Goldfinger, 
 E. D. Grundstrom\altaffilmark{1}, 
 W. Huang\altaffilmark{2},
 H. A. McAlister, 
 A. M\'{e}rand, 
 J. Sturmann, 
 L. Sturmann, 
 Y. Touhami, 
 N. H. Turner,
 D. W. Wingert\altaffilmark{1}}

\affil{Center for High Angular Resolution Astronomy and 
 Department of Physics and Astronomy,\\
 Georgia State University, P. O. Box 4106, Atlanta, GA 30302-4106} 
\email{gies@chara.gsu.edu, bagnuolo@chara.gsu.edu, baines@chara.gsu.edu, 
 theo@chara-array.org, farrington@chara.gsu.edu, pj@chara-array.org, 
 erika@chara.gsu.edu, wenjin@astro.caltech.edu, hal@chara.gsu.edu, 
 antoine@chara-array.org, judit@chara-array.org, sturmann@chara-array.org, 
 yamina@chara.gsu.edu, nils@chara-array.org, wingert@chara.gsu.edu}

\altaffiltext{1}{Visiting Astronomer, Kitt Peak National Observatory,
 National Optical Astronomy Observatory, operated by the Association
 of Universities for Research in Astronomy, Inc., under contract with
 the National Science Foundation.}

\altaffiltext{3}{Current address: Department of Astronomy, 
 California Institute of Technology, MS 105-24, 
 Pasadena, CA 91125}

\author{D. H. Berger} 
\affil{Department of Astronomy, University of Michigan,
 500 Church Street, 917 Dennison Building, Ann Arbor, MI 48109-1042}
\email{dhberger@umich.edu}

\author{M. V. McSwain\altaffilmark{3}}
\affil{Astronomy Department, Yale University, New Haven, CT 06520-8101}
\email{mcswain@astro.yale.edu}

\altaffiltext{3}{NSF Astronomy and Astrophysics Postdoctoral Fellow}

\author{J. P. Aufdenberg\altaffilmark{4}, S. T. Ridgway}
\affil{Kitt Peak National Observatory, National Optical Astronomy Observatory, \\
 P.O. Box 26732, Tucson, AZ 85726-6732}
\email{aufded93@erau.edu, sridgway@noao.edu}

\altaffiltext{4}{Current address: 
 Physical Sciences Department, 
 Embry-Riddle Aeronautical University,
 600 S.\ Clyde Morris Blvd., 
 Daytona Beach, FL 32114}

\author{A. L. Cochran, D. F. Lester, N. C. Sterling}
\affil{Department of Astronomy, University of Texas,  
 1 University Station, C1400, Austin, TX 78712} 
\email{anita@barolo.as.utexas.edu, dfl@astro.as.utexas.edu, 
 sterling@astro.as.utexas.edu} 

\author{J. E. Bjorkman, K. S. Bjorkman} 
\affil{Ritter Observatory, M. S. 113, Department of Physics and Astronomy,
 University of Toledo, Toledo, OH 43606-3390}
\email{jon@physics.utoledo.edu, karen@astro.utoledo.edu}

\author{P. Koubsk\'{y}}
\affil{Astronomical Institute, Academy of Sciences, Fri\v{c}ova 296,
 CZ-251 65 Ond\v{r}ejov, Czech Republic}
\email{koubsky@pleione.asu.cas.cz}

\slugcomment{ApJ, iin press (2007 Jan.\ 1 issue)}
\paperid{65639}


\begin{abstract}

We present the first $K^\prime$-band, long-baseline interferometric 
observations of the northern Be stars $\gamma$~Cas, $\phi$~Per, 
$\zeta$~Tau, and $\kappa$~Dra.  The measurements were made with 
multiple telescope pairs of the CHARA Array interferometer, 
and in every case the observations indicate that the 
circumstellar disks of the targets are resolved.   
We fit the interferometric visibilities with predictions from 
a simple disk model that assumes an isothermal gas in 
Keplerian rotation.  We derive fits of the four model parameters
(disk base density, radial density exponent, disk normal inclination, 
and position angle) for each of the targets.  The resulting 
densities are in broad agreement with prior studies of the IR excess 
flux and the resulting orientations generally agree with those
from interferometric H$\alpha$ and continuum polarimetric observations. 
We find that the angular size of the $K^\prime$ disk emission is 
smaller than that determined for the H$\alpha$ emission, 
and we argue that the difference is the result of a larger 
H$\alpha$ opacity and the relatively larger neutral hydrogen 
fraction with increasing disk radius.  All the targets are known 
binaries with faint companions, and we find that companions 
appear to influence the interferometric visibilities in the 
cases of $\phi$~Per and $\kappa$~Dra.  We also present contemporaneous
observations of the H$\alpha$, H$\gamma$, and Br$\gamma$ emission lines. 
Synthetic model profiles of these lines that are based on the 
same disk inclination and radial density exponent as derived from 
the CHARA Array observations match the observed emission line 
strength if the disk base density is reduced by $\approx 1.7$ dex. 

\end{abstract}



\setcounter{footnote}{4}

\section{Introduction}                              

Be stars are rapidly rotating B-type stars that manage to eject gas
into a circumstellar disk (observed in H emission lines, an IR excess 
flux, and linear polarization; \citealt{por03}).  Be star disks are 
ephemeral and vary on timescales from days to decades (in some cases
disappearing altogether for extended periods).   This inherent 
time variability suggests that gas injection into the disk is only 
partially the result of the fast spin and equatorial extension of the
Be star \citep*{tow04} and that magnetic, pulsational, wind driving, 
and/or other processes are required for mass loss into the disk  
\citep{owo05}.  The evolutionary status of Be stars is still a 
subject of considerable debate \citep*{mcs05,zor05}, but there is
growing evidence that many Be stars were spun up through mass 
transfer from a binary companion that has since become a neutron star,
white dwarf, or hot subdwarf \citep{pol91,gie00}.

The direct resolution of Be star disks has come
through high resolution radio observations ($\psi$~Per;  
\citealt{dou92}) and long baseline interferometry in a
narrow band around the H$\alpha$ emission feature \citep{ste05}. 
The first H$\alpha$ studies focused on the bright, northern sky  
Be stars $\gamma$~Cas \citep*{tho86,mou89,ste95,ber99} and $\zeta$~Tau
\citep{vak98}.  In a seminal paper, \citet{qui97} presented 
results for seven Be stars based upon narrow-band H$\alpha$ 
observations made with the MkIII interferometer at Mount Wilson 
Observatory.  They were able to resolve the disks around 
all seven stars, and they found diameters in the range $2.6 -
4.5$~mas (Gaussian FWHM).  In most cases the
visibilities indicated an elongated shape; for example, in the case of
$\zeta$~Tau, the major axis was resolved (4.53~mas diameter) but the
orthogonal minor axis remained unresolved, consistent with the
suggestion that we are viewing the disk almost edge-on.  The disk
orientation position angle on the sky inferred from polarimetry matched
the interferometric results in every case.  These results have been 
extended and improved through new interferometric observations 
with the Navy Prototype Optical Interferometer (NPOI) in an 
important series of papers by Tycner and collaborators 
\citep{tyc03,tyc04,tyc05,tyc06}.  
 
The relative flux contribution from the disk becomes larger 
in the infrared spectral range due to bound-free and free-free 
emission from the dense and ionized gas in the disk.  
\citet{wat86} developed a disk model for the IR emission 
that he subsequently applied to determine the disk properties of 101  
Be stars based upon their far-IR emission observed with 
the {\it IRAS} satellite \citep*{cot87,wat87}, and \citet{dou94} further 
extended this analysis to study the near-IR excess in 144 Be stars.   
\citet{how01} demonstrated that the near-IR flux excess is 
correlated with the H$\alpha$ emission strength in Be stars 
as expected for a common origin in the disk.  Thus, the near-
and mid-IR ranges \citep*{rin99} offer a new opportunity for resolving the disks 
of Be stars through long baseline interferometry.  \citet{ste01} presented 
a wind model for the infrared emission from Be stars and the corresponding 
interferometric visibilities for the models.  Their model for 
$\gamma$~Cas, for example, predicts that the disk should appear approximately 
two times larger in the $K$-band continuum and in the Br$\gamma$ emission 
line than found by narrow-band H$\alpha$ interferometry.  \citet{ste03} 
made additional predictions for 16 Be stars of the expected $K$-band 
interferometric visibilities.   The first attempt to resolve the
infrared disk flux was recently made by \citet{che05} who made $N$-band 
observations with the VLTI/MIDI instrument of the bright southern Be star 
$\alpha$~Ara, and their results indicate that the disk is smaller than 
that predicted by \citet{ste03}, perhaps due to disk truncation by a 
faint binary companion.   $K$-band interferometric observations from 
VLTI/VINCI of $\alpha$~Ara were reported by \citet{dom03} who 
interpreted the visibilities using a rotationally distorted photospheric model. 

Here we present the first near-IR interferometric observations 
of four bright, northern sky Be stars that we obtained with 
the Georgia State University Center for High Angular Resolution 
Astronomy (CHARA) Array at Mount Wilson Observatory \citep{mca05,ten05}.
We describe the interferometric and complementary spectroscopic 
observations in \S2 and then we present a simple disk model in \S3 
that we use to predict the interferometric visibilities and 
spectral line profiles.   We discuss the specific results for 
each of the four targets in \S4.  Our results are summarized in \S5.  


\section{Observations}                              

\subsection{Targets}

We selected four well known Be stars for this initial 
program of CHARA Array interferometry: 
$\gamma$~Cas (HR~264 = HD~ 5394 = HIP~4427),
$\phi$~Per (HR~496 = HD~10516 = HIP~8068),
$\zeta$~Tau (HR~1910 = HD~37202 = HIP~26451), and
$\kappa$~Dra (HR~4787= HD~109387 = HIP~61281).
All the targets except $\kappa$~Dra have prior 
narrow-band H$\alpha$ interferometric observations 
\citep{qui97,tyc03,tyc04,tyc06}.   
The first three stars were targets in speckle 
interferometric searches for close companions, 
and no such companions were detected \citep{mas97}; 
however, oscillations in the the proper motion 
of $\gamma$~Cas suggest the presence of a faint companion 
with a period of some 60~y \citep*{gon00}.  
On the other hand, all four of the targets are 
known spectroscopic binaries with periods in the 
range 61 -- 204~d.  The companion in the case of 
$\phi$~Per is a hot subdwarf whose spectral features 
appear in the UV spectrum \citep{gie98}, but for the 
other three stars the nature of the low mass companion is 
unknown.   

We summarize the adopted parameters for the
target stars in Table~1 that lists the spectral classification, 
parallax \citep{per97}, radius, mass, effective temperature, 
and projected rotational velocity \citep*{yan90,abt02} of the 
bright primary star.  The next three rows give the  
radius, mass, and effective temperature of the secondary. 
These are only known for $\phi$~Per and representative 
values are listed for the others assuming that they also
have hot subdwarf companions (see \S3.3).  The next 
four rows list the binary period, epoch of the secondary 
star's maximum radial velocity (equal to the epoch of the 
secondary's crossing of the ascending node), the 
primary star to center of mass portion of the semimajor axis, 
and the adopted Roche radius of the primary star. 
The final row provides a key to the references from which 
these various parameters were adopted.  

\placetable{tab1}      

\subsection{$K^\prime$-band Interferometry from the CHARA Array}

The CHARA Array observations were made on various dates during 
the first two years of operation (Table~2).  The telescope, 
instrumentation, and data reduction procedures are described in detail 
by \citet{ten05}.  The CHARA Array consists of six 1~m telescopes 
in a $Y$-configuration with pairwise baselines ranging from 
34 to 330~m in length.  The Be star observations were primarily 
made with intermediate to long baseline configurations using the 
$K^\prime$ filter described by \citet{mca05}.  A summary of 
each night's observations is given in Table~2 that lists 
the target and calibrator star names, the telescope pair (with 
the maximum baseline given in meters in parentheses; \citealt{ten05}), 
the UT date, and the number of sets of measurements.   
We made the measurements using the ``CHARA Classic'' beam combiner, 
which is a two-beam, pupil-plane (or Michelson) combiner \citep{stu03}.
The fringes were recorded with a near-IR detector on either 1 or 
$2\times2$ pixels at a sampling rate of 100 or 150 Hz depending on 
the seeing conditions.  The path length scans were adjusted to obtain 
approximately five samples per fringe spacing.  Each measurement set 
consists of some 200 scans of the fringe pattern with photometric 
calibration scans performed before and after.  The raw visibilities 
were estimated from an edited set of scans using the log-normal 
power spectrum method described by \citet{ten05}. 

\placetable{tab2}      

We transformed the raw visibilities into absolute visibility $V$ by 
interleaving the target observations with measurements of calibrator
stars with small angular diameters \citep{ber06}.  The calibrator stars
are generally close in the sky and of comparable $K^\prime$ magnitude
to the targets, and we list in Table~3 their effective temperatures 
and gravities as derived by other investigators (identified in column 7).  
We estimated the angular diameters of the calibrator stars by 
comparing their observed and model flux distributions.  
The angular diameter of the limb darkened disk $\theta_{LD}$ 
(in units of radians) is found by the inverse-square law:
\begin{equation}
{ {f_\lambda ({\rm observed})}\over
  {F_\lambda ({\rm emitted}) ~10^{-0.4 A_\lambda}} } =
  ( {R_\star / d} )^2 =
  {1\over 4} \theta_{LD}^2
\end{equation}
where the ratio of the observed and emitted fluxes
(reduced by the effects of interstellar extinction $A_\lambda$)
depends on the square of the ratio of stellar radius $R_\star$
to distance $d$.   Most of the calibrator stars are 
relatively close by and the interstellar extinction is 
negligible.  The one exception is HD~107193, and in this case 
we adopted an extinction curve law from \citet{fit99}
that is a function of the reddening $E(B-V)$ and the ratio of
total-to-selective extinction $R=A_V/E(B-V)$ (set at a value of 3.1).   

\placetable{tab3}      

The model fluxes were interpolated from the grid of models from 
R.\ L.\ Kurucz\footnote{http://kurucz.cfa.harvard.edu/} 
based upon the values of $T_{\rm eff}$ and $\log g$ in Table~3.
These model fluxes are based upon solar abundance, plane-parallel, 
local thermodynamic equilibrium (LTE),
line blanketed atmospheres with a microturbulent velocity of 
4 km~s$^{-1}$ that should be adequate for these
A- and F-type main sequence stars.   
We compiled data on the observed fluxes from the optical to the near-IR.
We found Johnson $UBV$ photometry for all the calibrators,
and these magnitudes were transformed to fluxes using the calibration
of \citet*{col96}.  We also included Str\"{om}gren photometry for
several calibrators using the calibration of \citet{gra98}.
Spectrophotometry from \citet*{kha88} was used for HD~6210. 
All of the stars are included in the {\it 2MASS All-Sky Catalog of Point Sources}
\citep{cut03}, and we converted the $JHK$ magnitudes to fluxes
using the calibration of \citet*{coh03}.  In a few cases the 
errors in the {\it 2MASS} magnitudes were unacceptably large, and 
we relied instead on IR fluxes from the {\it COBE} DIRBE instrument
\citep*{smi04}.  The angular diameters we derived by fitting the 
spectral energy distributions are given in column~6 of Table~3.  
Most of these calibrators have angular diameters that are small 
enough that their associated errors will introduce only minor
systematic errors in our derived target angular sizes \citep{van05}.

It is normal practice in interferometry to estimate the 
actual visibility of a calibrator by adopting a uniform disk 
function for the visibility curve as a function of wavelength 
and baseline, but in fact it is no more difficult to apply 
a visibility curve for a limb darkened disk provided the 
form of the limb darkening is known \citep*{dav00}.  
We used the effective temperatures and gravities from 
Table~3 to estimate the $K$-band limb darkening coefficients 
from the tables of \citet{cla00}, and then we calculated 
the limb darkened visibility curves for the calibrators 
(for the projected baseline at the time of observation
and the limb darkened angular diameter given in Table~3)
by making a weighted sum of the predicted visibilities 
over the wavelength band of the $K^\prime$ filter 
(to account for bandwidth smearing).   We then made 
a simple linear interpolation in time between the target
and calibrator observations to estimate the ratio of raw to 
absolute visibility required to transform the results to 
absolute visibility.  Note that we also made a spline fit of 
the time evolution of calibrator visibility, and the difference 
between the spline and linear interpolated visibilities was added 
quadratically to the error budget for the final calibrated visibility
to help estimate the errors introduced by the time interpolation 
scheme. 
 
The calibrated visibilities are presented in Table~4 (available 
in full in the electronic version of the paper).  The columns in this
table give the target name, the heliocentric Julian date of the 
mid-point of the data set, the binary orbital phase determined 
from the period and epoch given in Table~1, the telescope pair 
used for the observation, the projected baseline and position angle
of the target as viewed from the telescope pair at the mid-point time, 
and the calibrated visibility.  Note that the visibility errors 
quoted represent the internal errors from the power spectrum 
analysis of the fringes plus a term introduced for the calibration 
process, and these may underestimate the actual visibility errors. 
We find, for example, that there are some closely spaced observations 
on certain nights with comparable projected baselines and position
angles that have a scatter that is several times larger than the 
formal errors.  There are 11 such subsets of 4 or more measurements 
from within a specific night that have ranges of $\delta ({\rm baseline}) < 10$~m 
and $\delta ({\rm position~angle}) < 10^\circ$, and the average ratio of the 
standard deviation of $V$ within a subset to the mean of the 
quoted internal error in $V$ is a factor of 2.8. 
Thus, the quoted errors are probably lower limit estimates of 
the true error budget. 

\placetable{tab4}      

\subsection{Optical Spectroscopy}

The disks of Be stars exhibit large temporal variations in their
H-Balmer emission strengths that presumably reflect large 
structural changes in their geometry and/or density \citep{por03}. 
Thus, we obtained new spectroscopic observations of the 
target Be stars in order to compare the Balmer emission 
levels at times contemporaneous with the interferometric 
observations.   The sources and details of the spectroscopy
are listed in Table~5 that reports the observatory and 
telescope of origin, the spectroscopic instrument, 
spectral resolving power $R=\lambda/\delta\lambda$, 
spectral range recorded, UT dates, and the names of 
targets.  Most of the observations of the H$\alpha$ and 
H$\gamma$ lines were made with the Kitt Peak National 
Observatory (KPNO) coud\'{e} feed telescope, and these have moderate 
resolution and high S/N properties.   The H$\alpha$ spectra
of $\kappa$~Dra were obtained at both the University of Toledo 
Ritter Observatory and the Ond\v{r}ejov Observatory at the 
same time as the first CHARA Array observations were made,
and a low dispersion blue spectrum was also obtained at that
time with the University of Texas McDonald Observatory 2.7~m 
telescope (recording H$\beta$ through the higher Balmer sequence). 
All of these spectra were reduced using standard routines in 
IRAF\footnote{IRAF is distributed by the National Optical Astronomical
Observatory, which is operated by the Association of Universities for
Research in Astronomy, Inc. (AURA), under cooperative agreement with the
National Science Foundation.} to create continuum rectified
versions of each spectrum.  The atmospheric telluric lines in 
the vicinity of H$\alpha$ were removed from the KPNO coud\'{e} feed 
spectra \citep{gie02}, but these features were quite weak in the 
$\kappa$~Dra H$\alpha$ spectra and were left in place.  We describe
these Balmer line profiles in \S4 and \S5 below. 

\placetable{tab5}      

\subsection{$K$-band Spectroscopy}

The IR excess flux from Be star disks will tend to dilute the 
photospheric and emission lines in the near-IR spectral range, 
and the strength of the Br$\gamma$ emission line in particular 
offers a sensitive test of the IR excess fluxes derived from 
interferometry.   Thus, we also obtained low 
resolution $K$-band spectroscopy of all the targets except 
$\kappa$~Dra with the University of Texas McDonald Observatory 2.7~m 
telescope and CoolSpec spectrometer \citep{les00}. 
We used a $1\farcs0$ entrance slit and a grating with 
75 grooves mm$^{-1}$ in second order, yielding a spectral 
resolving power of $R\approx 1520$ (Table~5). 
The detector was a NICMOS3 HgCdTe $256\times256$ pixel array. 
The spectra were reduced using the methods described by \citet{lik06}
except that in our case the sky background was estimated from 
the intensity at off-star positions along the projected slit. 

We also made a number of spectra at different air masses of the 
broad-lined, A0~V stars HR~567 (HD~11946) and HR~2398 (HD~46553), 
which we used to remove the numerous atmospheric features in this 
spectral band.  The stellar spectra of these two stars are essentially 
featureless with the exception of the photospheric Br$\gamma$ 
line \citep{wal97}, and we formed pure atmospheric spectra by 
dividing our observed spectra by the average stellar spectra 
of the similar stars HR~5793 and HR~7001 (after broadening 
for the lower resolution of our spectra and for the higher 
projected rotational velocities of HR~567 and HR~2398) 
that we obtained from the spectral library of 
\citet{wal97}\footnote{ftp://ftp.noao.edu/catalogs/medresIR/K\_band/}.
The target spectra were then divided by these atmospheric 
spectra (after small line depth corrections to account for 
differing air masses) to obtain the final continuum rectified 
versions that are presented below (\S5).


\section{Be Star Disk Models}                       

The observed interferometric visibilities can be interpreted by 
assuming that the Be star disks have a relatively simple 
spatial intensity in the sky, perhaps a uniform ellipsoid,  
ring, or Gaussian ellipsoid \citep{tyc06}. 
However, the quality of the interferometric data is sufficient 
to begin to explore more physically motivated models for the 
disk appearance.  Here we present a relatively simple model 
for the near-IR emission that is derived from the Be 
circumstellar disk model of \citet{hum00}.  The advantages of 
this approach are that we can compare the resulting disk gas 
densities with prior work on the near-IR spectrum and 
that we can also predict the disk emission line strengths 
for comparison with our observed spectra.  In the following 
subsections we outline the basic assumptions and characteristics
of the model that we then apply to fit the CHARA Array 
observations of visibility.  We discuss the model predictions
for spectroscopy below in \S4. 

\subsection{Disk Geometry and Model Parameters}

\citet{hum00} present a simple model for the circumstellar 
disks of Be stars that they use to explore the emission line 
shapes.   The basic concept is that the disk is axisymmetric 
and begins at the stellar surface.  The gas density in the 
disk is given by 
\begin{equation}
\rho (R,Z) = \rho_0 R^{-n} \exp \left[-{1\over2}\left({Z\over{H(R)}}\right)^2\right]
\end{equation}
where $R$ and $Z$ are the radial and vertical cylindrical 
coordinates (in units of stellar radii), $\rho_0$ is the base 
density at the stellar equator, $n$ is a radial density 
exponent, and $H(R)$ is the disk vertical scale height.  
This gas scale height is given by 
\begin{equation}
H(R)={C_S\over V_K} R^{3\over2}
\end{equation}
where $C_S$ is the sound speed and $V_K$ is the 
Keplerian velocity at the stellar equator.  We assume that 
the outer boundary of the disk occurs at a radius 
$R_d$ that is equal to the Roche radius of the Be star 
for these four binary targets (Table~1).  This assumption 
is probably more critical for the mid-IR emission spectrum, 
but it not important for the $K^\prime$ emission that is generally 
confined to regions well within the Roche radii of our targets. 

\citet{hum00} assume that the temperature is constant throughout
the disk and is given by $T_d = {2\over3} T_{\rm eff}$ where 
$T_{\rm eff}$ is the stellar effective temperature.  
In fact, recent physical models for Be disks by \citet{car06} 
show that the gas temperature is actually a function of 
both $R$ and $Z$, but their work shows that the temperature is
about $60\% ~T_{\rm eff}$ in the outer, optically thin
parts of the disk, so we adopt this value for the isothermal 
approximation used here.   

We assume that the star itself is spherical and of uniform 
brightness in the $K^\prime$ band.  Neither of these assumptions
is probably correct for a rotationally distorted and gravity 
darkened Be star \citep{tow04}, but the photospheres of the 
targets are small enough (see Table~6 below) that these details are
inconsequential for the baselines and wavelength of the 
CHARA Array observations.  We suppose that the disk is observed 
with the disk normal oriented to our line of sight with 
an inclination angle $i$ and a position angle $\alpha$ 
measured east from north in the sky. Note that this position 
angle convention is $90^\circ$ different from that adopted
in the H$\alpha$ interferometric studies where the 
position angle of the projected major axis is usually given. 
Thus, the four parameters that define a disk model are 
the base density $\rho_0$, the density exponent $n$, 
and the orientation angles $i$ and $\alpha$.

\subsection{$K^\prime$-band Continuum Images and Interferometric Visibility}

We determine the surface brightness of the disk plus star over 
a projected rectilinear coordinate grid on the sky 
by solving the equation of transfer 
along a ray through the center of each grid position, 
\begin{equation}
I = S_{d} (1 - e^{-\tau}) + 
            I_{\star} e^{-\tau}
\end{equation}
where $I$ is the derived specific intensity, 
$S_{d}$ is the source function for the disk gas 
(taken as the Planck function for the disk temperature $T_d$), 
$I_{\star}$ is the specific intensity for a uniform 
disk star (taken as the Planck function for $T_{\rm eff}$), 
and $\tau$ is the integrated optical depth along
the ray.  The main shortcoming in this expression is 
the neglect of a scattered light term to account for 
Thomson scattering of photospheric intensity
(see, for example, eq.\ [6] in \citealt{bjo94}). 
For the typical electron densities that we find for 
our Be star targets, this term will amount to only a 
few percent of the stellar specific intensity at 
locations close to the star, which is usually much less 
than the disk source function in the optical thick portions, 
so this simplification is acceptable in the $K^\prime$ band. 

The disk optical depth in the near-IR is predominantly 
due to bound-free and free-free processes, and we can express 
the increment in optical depth with an incremental step 
$ds$ along a given ray as 
\begin{equation}
d\tau = C(\lambda, T_d) \rho (R,Z)^2 ~ds
\end{equation}
where the coefficient $C(\lambda, T_d)$ is given by 
equation (5) in \citet{dou94} (see also \citealt{wat86}).
This coefficient includes terms for the Gaunt factors for 
bound-free and free-free emission that we evaluated for the 
$K^\prime$ band using the tables in \citet{wat84}. 
We adopted an ionization model for the disks of the two 
hotter Be stars, $\gamma$~Cas and $\phi$~Per, that 
assumed ionized H, singly-ionized He, and doubly-ionized C, 
N, and O atoms, while for the cooler Be stars, $\zeta$~Tau 
and $\kappa$~Dra, we assumed ionized H, neutral He, and 
singly ionized C, N, and O.  Note that we evaluated the 
optical depth coefficient only at the central wavelength 
of the $K^\prime$ filter since the coefficient varies 
slowly with wavelength.  Also note that 
we have accounted only for continuum emission in this band 
since the Br$\gamma$ emission contribution is small compared 
to the flux integrated over the $K^\prime$ band (\S5).  

The image of the Be star in the sky in constructed by 
integrating the optical depth along each ray according to 
equations (2) and (5), and then populating each pixel of 
the image using equation (4).  The pixel scale on the sky is 
set by the adopted angular diameter of the star (derived 
from the parallax and radius in Table~1).  We show an 
example calculation for $\gamma$~Cas in the left panel of
Figure~1.  Portions of the disk projected against the sky 
include only the first term of equation (4), and the 
disk intensity attains a maximum of $S_{d} \approx 55\% I_{\star}$
in the inner optically thick regions.  At a radius where the 
optical depth is approximately unity, the disk becomes optically 
thin and fades with increasing radius.  Part of the 
background disk is occulted by the star while the part in 
foreground attenuates the photospheric intensity.   
We also include a binary star option to add the disk of 
a secondary into the image at a position appropriate for 
the time of observation. 

\begin{figure}
\begin{center}
{\includegraphics[angle=90,height=10cm]{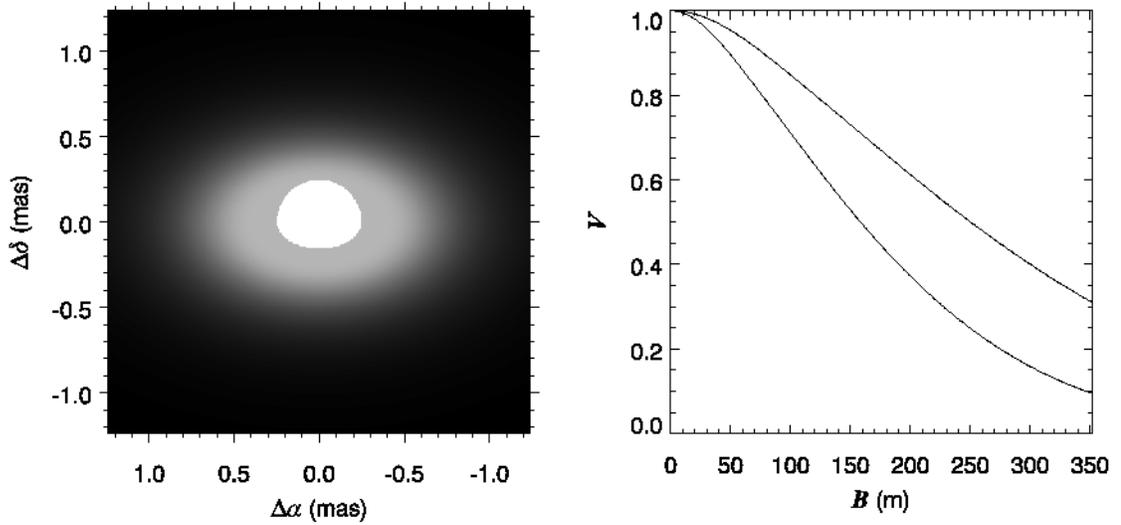}}
\end{center}
\caption{The left panel is a depiction of the $K^\prime$ 
image of a disk model for the case of $\gamma$~Cas  
($i=51^\circ$, $\rho_0 = 7.2\times 10^{-11}$ g~cm$^{-3}$,
$n=2.7$).  The right panel shows the corresponding 
interferometric visibility $V$ as a function of projected 
baseline for position angles along the apparent minor axis
({\it top line}) and major axis ({\it bottom line}).
}
\label{fig1}
\end{figure}

\placefigure{fig1}     

The interferometric visibilities are then calculated by 
making Fourier transforms of such disk images, and we did this 
following the techniques described by \citet{auf06}. 
The comparison with the observed visibility is done by 
determining the discrete Fourier transform of the 
image for the $(u,v)$ spatial frequency pair of the 
observation (see eq.\ [19] in \citealt{auf06}), 
and this calculation is done successively and 
summed over five wavelength bins according to the 
transmission curve of the $K^\prime$ filter \citep{mca05}. 
This procedure assumes that spectral energy distributions
in the $K^\prime$ band of our targets are similar to 
the stellar transmission curve (Rayleigh-Jeans tail), 
and this is probably acceptable in the $K^\prime$ band
where the disk emission is just beginning to become an
important component of the total flux \citep{wat86}. 
The Fourier transforms along the major and minor 
axes of the projected disk image for the $\gamma$~Cas example
are illustrated in the right hand panel of Figure~1 (here 
without accounting for bandwidth smearing).   The top curve shows 
the slower decline in $V$ for the smaller minor axis while the 
bottom curve illustrates the rapid decline for the better 
resolved major axis.   Note that the star itself is so 
small ($\theta_{LD}=0.50$ mas) that its visibility curve 
in the absence of a disk would decline to only $V=0.87$ 
at a baseline of 300~m. 

Figure 2 shows how the image and visibilities change by 
altering the disk inclination angle from $51^\circ$ to 
$80^\circ$ (the dotted lines in the right hand panel copy 
the original visibility curves from Fig.~1).  
The rays through the outer positions along the major 
axis now traverse more optical depth because of the 
oblique view through the disk, and this brightens the 
outer regions making the disk effectively larger in this
dimension (so that the decline in $V$ with baseline 
is steeper).  On the other hand, the minor axis now
appears more foreshortened and the visibility decline 
is reduced.  Thus, the $V$ variation with position 
angle in the sky (with respect to the major axis) 
is very sensitive to the disk inclination angle $i$. 

\begin{figure}
\begin{center}
{\includegraphics[angle=90,height=10cm]{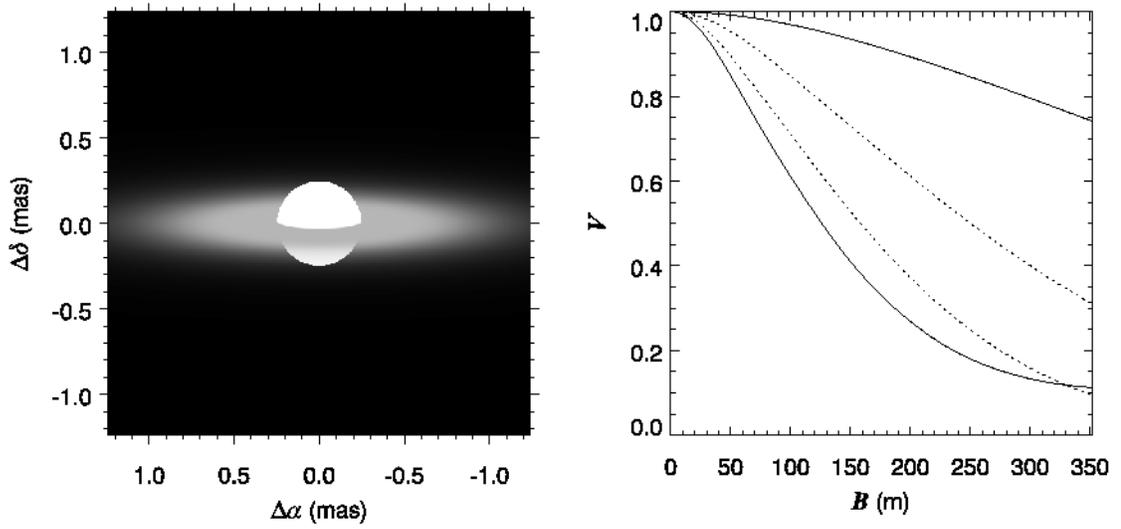}}
\end{center}
\caption{An image and visibility plots for a $\gamma$~Cas 
model as in Fig.~1 but with a larger inclination of $i=80^\circ$.
The dotted lines in the visibility panel repeat the curves from
the original model in Fig.~1.}
\label{fig2}
\end{figure}

\placefigure{fig2}     

The changes resulting from a decrease in disk base density  
$\rho_0$ are illustrated in Figure~3.  Now the density is lower
throughout the disk, and consequently the optically thick part of
the disk appears smaller in both dimensions (and the $V$ curves
decline more slowly with baseline).  Changes in the opposite 
sense result from a decrease in the density exponent as shown 
in Figure~4 (from $n=2.7$ to 2.0).   The radial density decline 
is less steep in this case so that much more of the inner disk 
is optically thick, leading to a larger appearance in the sky 
(steeper decline in the $V$ curves).  These examples show that 
the visibility curves are most sensitive to the disk size as
given by the boundary between the optically thick and thin 
regions.  Since the optical depth unity radius in the disk plane 
depends on $\rho_0^2 r^{-2n+3/2}$, there will be a locus of 
parameters that will produce visibility 
curves with similar decline rates in the main lobe and with 
subtle differences appearing only at intermediate and longer baselines.  
Thus, accurate determinations of both $\rho_0$ and $n$ will 
require visibility measurements over a broad range of baselines. 

\begin{figure}
\begin{center}
{\includegraphics[angle=90,height=10cm]{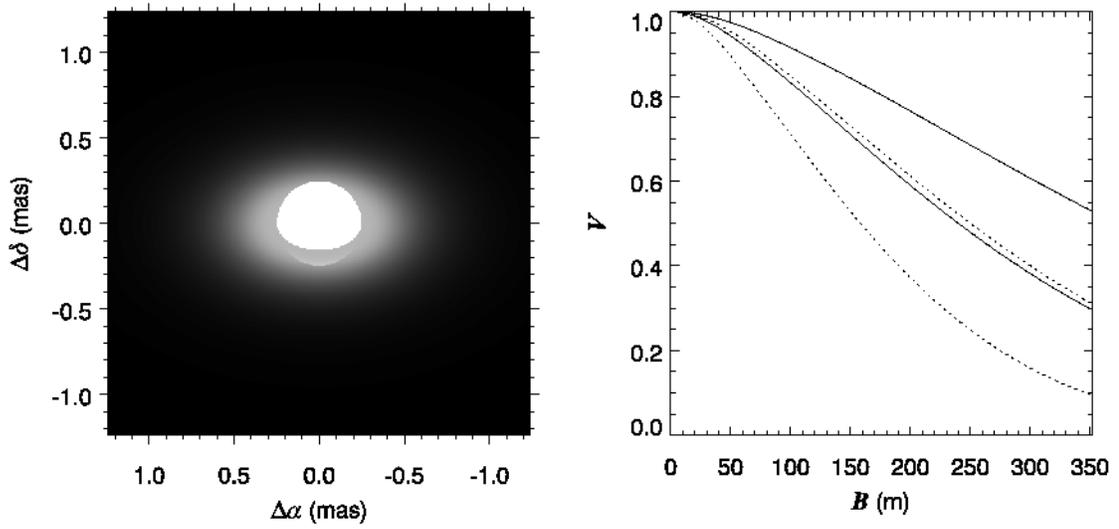}}
\end{center}
\caption{An image and visibility plots for a $\gamma$~Cas 
model as in Fig.~1 but with a smaller base density of 
$\rho_0=3.6\times 10^{-11}$ g~cm$^{-3}$.
The dotted lines in the visibility panel repeat the curves from
the original model in Fig.~1.}
\label{fig3}
\end{figure}

\placefigure{fig3}     

\begin{figure}
\begin{center}
{\includegraphics[angle=90,height=10cm]{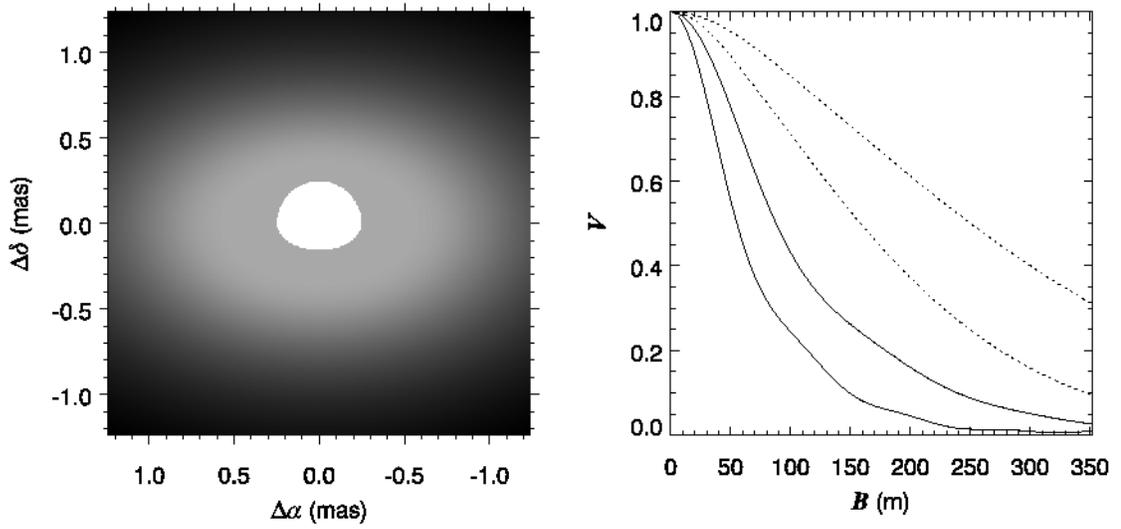}}
\end{center}
\caption{An image and visibility plots for a $\gamma$~Cas 
model as in Fig.~1 but with a smaller radial density 
exponent of $n=2.0$. 
The dotted lines in the visibility panel repeat the curves from
the original model in Fig.~1.}
\label{fig4}
\end{figure}

\placefigure{fig4}     

\subsection{Model Fitting Results}          

We found best fit solutions for the four disk parameters 
by fitting the model and observed $K^\prime$ visibilities 
using a numerical grid search technique.  We used an iterative 
search on each of the individual parameters to find the set
that provided the lowest overall $\chi^2_\nu$ statistic. 
The results are listed in Table~6 for both single star and 
binary star models (see below).  The seventh row of Table~6 
reports the minimum $\chi^2_\nu$ values, and these are all 
much larger than the expected value of unity for acceptable fits. 
Recall that the internal errors on $V$ may underestimate 
the actual errors by a factor of 2.8 for these measurements
(\S2.2), so we might instead expect to find best fits with 
$\chi^2_\nu \approx 2.8^2 = 7.84$.  This is more or less 
consistent with the fitting results for all but $\gamma$~Cas, 
where the residuals are still larger than expected. 
We estimated the errors in the 
fitting parameters by renormalizing the minimum $\chi^2_\nu$ 
to unity and then finding the excursion in the parameter 
that increased $\chi^2_\nu$ to a value of $1+1/(N_V-4)$ 
(for $N_V$ visibility measurements and 4 fitting parameters). 
Since the disk parameters $\rho_0$ and $n$ are so closely 
coupled in the fits (\S3.2), we searched the locus of the 
pair-wise fits for the same increase in $\chi^2_\nu$,  
and we quote the half-range in these parameters for their
errors. 

\placetable{tab6}      

The first two rows of Table~6 give the orientation parameters, 
the disk plane normal inclination $i$ and position angle $\alpha$ (east from north). 
Note that there is a $180^\circ$ ambiguity in the value of $\alpha$
(i.e., whether the disk north pole is in the eastern or western half 
of the sky), so we assume the range $0\leq \alpha < 180^\circ$. 
The third and fifth rows list the logarithm (base 10) of the disk base 
density $\log \rho_0$ and the disk density exponent $n$, respectively
(base electron density is $\log N_e = \log \rho_0 + 23.6$). 
These are compared with the same parameters in rows four and six, respectively, 
estimated from model fits of the IR flux excess from {\it IRAS} 
observations that were computed by \citet{wat87}.  \citet{wat87} 
make a number of simplifying assumptions about the disk geometry
that make a direct comparison of results unreliable, but in 
general there is reasonable agreement except in the two cases 
where a binary companion may influence the results, $\phi$~Per
and $\kappa$~Dra (see below).  Row 8 gives a derived model 
quantity related to the IR flux excess, 
$E(V-K) = 2.5\log (1+F_d/F_\star)$, 
where $F_d$ and $F_\star$ are the integrated disk and 
stellar flux, respectively.  If we assume that the disk flux 
is relatively small in the optical, then this quantity can be directly compared
to the observed IR flux excess measurements by \citet{dou94} 
that are listed in row 9.  The model and observed values of 
$E(V-K)$ are comparable but not always a good match.  
We suspect the differences may relate to the temporal variations 
in the disk emission (from  $\sim 1990$ for the IR magnitudes to 
$\sim2005$ for the CHARA Array measurements).  Rows 10 and 11 
compare the adopted stellar angular diameter $\theta_{LD}$ 
with the model disk emission FWHM diameter along the projected 
major axis.  The disk emission FWHM was estimated from the width 
at $50\%$ of the emission intensity in the optically thick part 
just above the photosphere. 

We know that all four targets are close binaries with faint, low-mass
companions, but the nature of the companion is only known in the 
case of $\phi$~Per \citep{gie98}.   The secondary of $\phi$~Per is
a hot helium star, the stripped down remains of what was once the 
more massive star before mass transfer began.  If we assume that 
all our Be star targets are likewise the mass gainers that have 
been spun up by the angular momentum carried with mass transfer, 
then they may also have helium star companions (or possibly 
white dwarf or neutron star companions).  We present here a sample 
calculation of how such a binary companion would affect the $K^\prime$ 
visibilities.  For the purpose of this demonstration, 
we assume that the companion is a hot 
($T_{\rm eff} = 30000$~K) and small ($R=1 R_\odot$) helium star
(except in the case of $\phi$~Per where we have better estimates for 
these parameters; see Table~1).   According to the interacting binary 
scenario, the angular momentum vector of the mass gainer Be star 
should be parallel with the orbital angular momentum vector,  
so the disk plane corresponds to the orbital plane.  
Consequently, we can apply the orientation parameters for
the disk, $i$ and $\alpha$, also to the orbit of the secondary. 
We know the epoch of the secondary's ascending node passage from 
the spectroscopic radial velocity curves (Table~1), so the only 
remaining unknown for the secondary's astrometric orbit is the 
sense of motion on the sky (increasing or decreasing position 
angle with time).  We follow the normal convention of assigning 
an inclination in the range $0^\circ\leq i < 90^\circ$ for counterclockwise 
apparent motion and in the range $90^\circ < i\leq 180^\circ$ for clockwise
motion (so that the longitude of the node is $\Omega = \alpha + 270^\circ$
for $i < 90^\circ$ and $\Omega = \alpha + 90^\circ$ for $i > 90^\circ$).  
We made fits of the visibility for these two binary orbit cases, 
and we show in Table~6 the results for the best fit sense of motion.
The minimum $\chi^2_\nu$ was significantly reduced only for the case of 
$\phi$~Per (where the properties of the secondary are known; the 
estimated magnitude difference is $\triangle K^\prime = 2.9$ mag).  The 
inclusion of the binary companion also resulted in a significant change in 
the fitted disk properties for $\kappa$~Dra (discussed further in \S5.4). 

It is valuable to compare the $K^\prime$ angular diameters with those
derived from narrow-band H$\alpha$ interferometry, and to do so we need 
to fit the visibilities with a simple Gaussian ellipsoidal model 
as done for the H$\alpha$ results.  We made these
fits also by a numerical grid search of the four parameters that 
describe the Gaussian ellipsoidal model: 
the angular FWHM of the major axis $\theta_{\rm GD}$, 
the ratio of the minor to major axes $r$ ($\geq \cos i$), 
the position angle of the major axis $\phi$ ($= \alpha \pm 90^\circ$), 
and the fraction of the total flux contributed by the 
photosphere of the Be star $c_{\rm P}$. 
The best fit results are listed in Table~7 for these single star models. 
Our results are compared with those from \citet{qui97} 
and \citet{tyc04,tyc06} based on interferometric
observations of the H$\alpha$ emission, and the orientation parameters 
appear to be in reasonable agreement.   We find, however, that the $K^\prime$ 
disk diameters are somewhat smaller than the H$\alpha$ diameters 
($\approx 64\%$ as large), contrary to the wind model predictions of \citet{ste01}.
Note that the $\chi_\nu^2$ residuals of the Gaussian elliptical fits 
are comparable to those from the physical model fits (Table~6), and  
additional longer baseline observations will be required to distinguish 
between them. 

\placetable{tab7}      


\section{Model Emission Lines and Spectroscopy}     

We can also use the simple disk model to estimate the 
line profiles of the disk emission lines \citep{hum00}. 
However, the problem is now more complex because of the 
wavelength dependence of the emission, and a number of 
key assumptions about the atomic populations must be made. 
Despite these difficulties, it is nevertheless interesting 
to check whether or not the disk density parameters 
derived from interferometry lead to predicted line profiles 
that are consistent with the observed ones.  Here we discuss 
the line synthesis predictions for the disk models and 
their comparison with our spectroscopic observations. 

\subsection{Disk Emission Line Models}

The line synthesis procedure has many similarities with 
the disk image calculations present above.  We assume the 
same basic density law (eq.\ [2]), the same creation of a 
rectilinear grid of points covering the projected positions 
of the star and disk in the sky, and the solution of the 
equation of transfer (eq.\ [4]) along along a ray through the
center of each grid point.  However, for the line synthesis  
the expression for the optical depth (eq.\ [5]) becomes \citep{hum00}
\begin{equation}
d\tau_\lambda = {{\pi e^2}\over{mc}} f \lambda_0 ~N(H_i) ~P_\lambda(V_r, N_e) ~ds
\end{equation}
and now we consider a wavelength dependent optical depth 
$\tau_\lambda$ for the region surrounding an emission line 
(we adopt an equivalent radial velocity grid in 10 km~s$^{-1}$ 
increments from $-2000$ to $+2000$ km~s$^{-1}$). 
The first term in the expression is the 
scattering coefficient for a classical oscillator 
that is multiplied by the oscillator strength 
$f$ and the rest wavelength $\lambda_0$ for the specific transition
(H$\gamma$, H$\alpha$, and Br$\gamma$). 

The next term $N(H_i)$ is the local number density of neutral 
H atoms in energy level $i$.   This can be set approximately by 
balancing the recombination rate with the photoionization rate 
(for the case when all the bound-bound transitions are very
optically thick; \citealt*{cas87})
\begin{equation}
\alpha (T) N_e N_p = N_i \int\limits_{\nu_0}^{\infty} {{4\pi}\over{h\nu}} a_\nu J_\nu ~d\nu
\end{equation}
where $\alpha (T)$ is the recombination coefficient, 
$N_e$ and $N_p$ are the electron and proton number densities, 
$N_i$ is the number density of hydrogen in energy level $i$, 
$a_\nu$ is the photoionization cross section, and $J_\nu$
is the mean radiation field.  If we assume that the latter 
results only from unattenuated starlight, then the mean 
field at a distance $r$ from the star can be written
\begin{equation}
J_\nu = W(r) I_\nu^\star 
\end{equation}
where $W(r)$ is the dilution factor given by 
\begin{equation}
W(r)=0.5(1-[1-r^{-2}]^{1\over2}), 
\end{equation}
$r$ is the radial distance from the star $r=(R^2+Z^2)^{1\over2}$,
and $I_\nu^\star$ is the stellar specific intensity.
\citet{cas87} rewrite the photoionization/recombination 
equilibrium to express the level population $N_i$ in terms of 
a variable $q_i$,
\begin{equation}
N_i=N_e N_p ~\alpha (T) / \int\limits_{\nu_0}^{\infty} {{4\pi}\over{h\nu}} a_\nu W(r) I_\nu^\star ~d\nu 
   \equiv N_e N_p q_i.
\end{equation}
In the stellar photosphere, $W(r)=1$, $I_\nu^\star = B_\nu (T_{\rm eff})$ (the Planck function),
and the gas will be in LTE, so that $q_i^{\rm LTE}=N_i/(N_e N_p)$ 
will be given by the Boltzmann-Saha equation.  In the circumstellar disk, the photoionization rate
will vary with $W(r)$ and the recombination rate will depend on $\alpha (T) \propto T^{-1/2}$,
so level populations will be given by 
\begin{equation}
N_i=q_i^{\rm LTE} \sqrt{T_{\rm eff}/T_d} ~N_e N_p / W(r).
\end{equation}
The actual populations probably have a more complex 
dependence on position in the disk \citep{car06}.  

The final term in optical depth equation is the normalized line
profile $P_\lambda(V_r, N_e)$ that depends on the 
local radial velocity $V_r$, electron density, and the adopted 
temperature of the isothermal disk $T_d$.  We adopted the convolved 
thermal and Stark broadened profiles calculated by T.\ Sch\"{o}ning and 
K.\ Butler\footnote{http://nova.astro.umd.edu/Synspec43/data/hydprf.dat}
for H$\alpha$ and H$\gamma$, but we used only a simple thermal profile 
for Br$\gamma$.  These profiles were collected into a matrix of
401 wavelength points (corresponding to our radial velocity grid), 
60 positive Doppler shifts (in increments of 10 km~s$^{-1}$), and
four electron number densities ($\log N_e = [10.5, 12, 13, 14]$). 

The line synthesis calculation proceeds by solving for the 
intensity spectrum from each pixel in the projected image and 
then summing these to create a predicted flux profile. 
At each position along a ray through a pixel, we determine the
Keplerian velocity and Doppler shift, gas density, H population, 
and profile shape (simply adopting the closest match in the grid), 
and then we integrate the optical depth for each wavelength point. 
The summed spectral intensity is then found through equation (4), 
but in this calculation $I_\star$ is replaced with a specific 
intensity corresponding to a photospheric absorption line with 
the corresponding value of surface normal orientation and 
Doppler shift for the projected position.  These intensity profiles
were calculated by first creating line blanketed LTE model atmospheres 
using the ATLAS9 program written by Robert Kurucz.  These models assume 
a stellar temperature and gravity derived from the parameters in Table~1, 
solar abundances, and a microturbulent velocity of 2 km~s$^{-1}$. 
Then a series of specific intensity spectra were calculated
using the code SYNSPEC \citep{hub95} for each of these models 
and H lines.  These profiles include the effects of limb darkening 
for the stellar photosphere.   Because our model assumes axial 
symmetry, we complete the calculation for only the half of the 
projected image with positive radial velocity and then we add 
a mirror-symmetric version of the flux profile to account for 
the other half with negative radial velocity.  The resulting 
profiles are then rectified to unit continuum flux and 
smoothed by convolution with a Gaussian function to match the 
instrumental broadening of our spectroscopic observations (Table~5).  
We also need to account for the disk IR continuum excess in the 
case of Br$\gamma$, so we renormalize the continuum flux $F_\lambda$ to 
\begin{equation}
F_\lambda^\prime = (F_\lambda + \epsilon) / (1 + \epsilon)
\end{equation}
where $\epsilon = 10^{0.4 E(V-K)} - 1$ ($E(V-K)$ given in Table~6). 

\subsection{Emission Line Fitting Results}

The model emission profiles depend on three parameters, 
the disk base density $\rho_0$, the density exponent $n$, and 
the disk inclination $i$.  Since our main interest is in 
checking the consistency of the interferometric and 
spectroscopic results, we set the values of $n$ and $i$ 
from the fits of the CHARA Array visibilities and then 
varied $\rho_0$ to find the best match with the observed 
line profiles.  Our results are gathered in Table~8 for 
each of the four targets.  The first three rows give the 
measured equivalent widths (where a negative sign indicates 
a net line flux above the continuum) for H$\alpha$, H$\gamma$, 
and Br$\gamma$, respectively.  The next three rows give the
corresponding disk base density $\rho_0$ that yields 
synthetic profiles with the observed equivalent width 
(or that best match the profile structure in the case 
of H$\gamma$ where the equivalent width is complicated by 
the presence of line blends).  The final row lists 
for comparison the value of $\rho_0$ as derived from the 
interferometric data (single star results for $\gamma$~Cas 
and $\zeta$~Tau; binary star results for $\phi$~Per and 
$\kappa$~Dra).   We see that there is relatively good agreement
between the densities derived from the three different 
H lines, and specifically the near agreement of the result from 
Br$\gamma$ with those from the Balmer lines supports 
the flux renormalization near Br$\gamma$ (eq.\ [12])
set by the interferometric determination of the IR excess.
The emission line and interferometric densities vary in tandem
among the analyses of the four stars, 
but we find that the emission line densities are 
systematically lower by approximately 1.7~dex.  
We suspect that the problem lies in the simplying assumptions
made in the model concerning the isothermal disk temperature 
and H populations that both probably lead to overestimates 
of the line emission from the outer parts of the disk.  

\placetable{tab8}      

The synthetic model profiles are directly compared with the 
observed ones in the next section, and we find that all the 
strong emission lines (formed in optically thick regions) 
appear to be much broader than predicted.  We show one example 
of this difference in Figure~5 for the H$\alpha$ line in 
$\gamma$~Cas.  The basic synthetic model that matches the 
observed equivalent width shows the strong double peaks 
associated with Keplerian motion in the disk.   The observed
profile, on the other hand, lacks well defined peaks and 
shows much broader line wings.  We also show in Figure~5 a 
smoothed version of the synthetic profile formed by convolution 
with a Voigt function fit using a damping parameter $\Gamma = 934$ 
km~s$^{-1}$, and this broadened version matches the observed 
profile quite well.  Thus, it appears that the line broadening 
due to Keplerian motion in the disk is insufficient to explain 
the shapes of the strong lines (although the weaker H$\gamma$ 
profiles are in better agreement; see \S5).  There are a number  
of possible explanations for the discrepancy: \\
(1) We have assumed that the H$\alpha$ source function is 
constant with radius in the disk, but more detailed models 
by \citet{car06} show that the temperature and presumably 
the source function decrease with radius.  This suggests that
the H$\alpha$ emission is probably stronger in the inner 
disk than portrayed in our model, and since large orbital 
motions in the inner disk produce the line wings, the model 
almost certainly underestimates the wing emission. \\ 
(2) Macroscopic motions such as stellar wind outflow \citep{ste95} 
and turbulence may be important (especially close to the star 
where the emission is particularly strong). \\
(3) Non-coherent scattering of photons into the wings of strong 
emission lines may extend their widths by as much as a few 
hundred km~s$^{-1}$ \citep{hum92,hum95}, but this may be insufficient 
to explain the extreme broadening observed in H$\alpha$. \\
(4) Thomson scattering by free electrons in the inner disk 
may create extended line wings \citep{poe79}. \\
(5) Collisional Stark broadening will produce similar extended 
wings, but tests with our model suggest that the base density 
would need to be increased by about two orders of magnitude for
the Stark wings to become prominent.  This might be possible if 
disk gas consists of dense clumps. \\
(6) Raman scattering of photons from the vicinity of the 
ultraviolet Ly$\beta$ line could lead to broad wings in H$\alpha$
\citep*{yoo02} if there is a sufficient column density of 
neutral hydrogen in the disk ($N({\rm H~I}) \sim 10^{20-23}$ cm$^{-2}$). 
This process might be especially applicable to edge-on 
and/or cooler disks.
 
\begin{figure}
\epsscale{1.}
\plotone{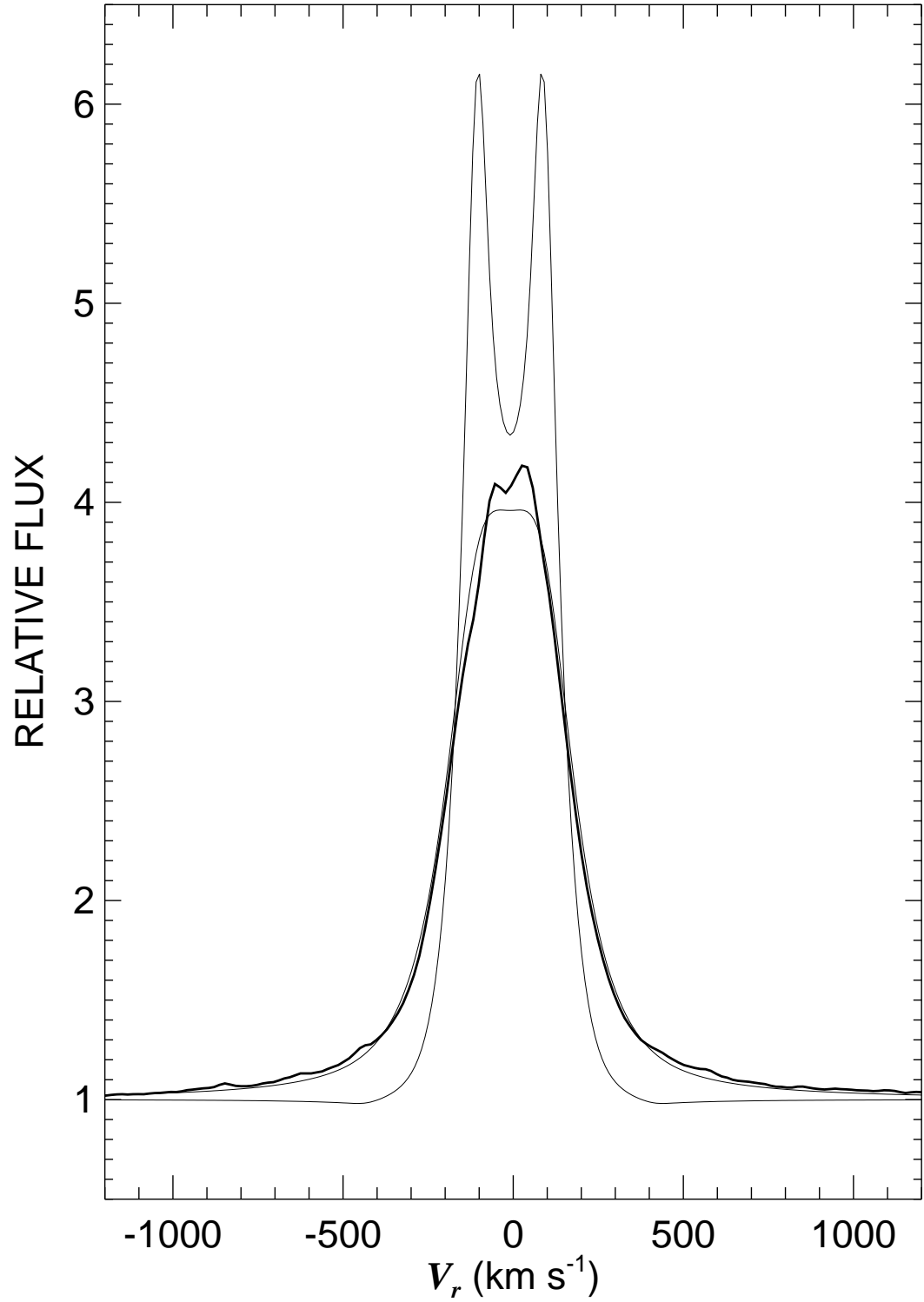}
\caption{The observed H$\alpha$ line in $\gamma$~Cas 
({\it thick line}) compared with the synthetic profile for
a Keplerian disk ({\it double-peaked thin line}) and the 
same convolved with a Voigt profile ({\it broader thin line}). }
\label{fig5}
\end{figure}

\placefigure{fig5}     


\section{Individual Targets}                        

\subsection{$\gamma$ Cas}

The star $\gamma$~Cas is perhaps the best documented Be star 
in the sky, and it has displayed Be, Be-shell, and normal B-type 
spectra over the course of the last century \citep{doa82}.  
It currently displays strong H$\alpha$ emission, and the 
H$\alpha$ disk size was recently determined through 
observations from NPOI \citep{tyc06}.  Our best fit disk (Table~6) 
and Gaussian elliptical models (Table~7) are generally consistent
with the H$\alpha$ interferometric results and indicate that 
we view this Be star disk at an intermediate inclination.  
The disk density exponent we derive, $n=2.7\pm0.3$, matches within 
errors the value $n=2.8\pm0.1$ found by \citet{hon00} based on an 
analysis of the 2.4 -- 45 $\mu$m spectrum from ISO-SWS.  We show 
the model $K^\prime$ image of $\gamma$~Cas in the left hand panel
of Figure~6 and the corresponding visibility image (without the 
effects of bandwidth smearing) is shown in the right hand panel. 
Both images are scaled logarithmic representations made to accentuate  
the low level features.   The $(u,v)$ frequencies of the CHARA Array 
observations are shown in the upper half of the visibility image 
with symbol sizes proportional to the observed visibility. 
They also appear in point-symmetric locations in the lower half 
where they represent the residual-to-error ratio (vertical segment
for $V_{\rm obs}>V_{\rm cal}$ and horizontal for $V_{\rm obs}<V_{\rm cal}$).
The residuals from the fit are large for our observations of $\gamma$~Cas, 
but they do not appear to indicate any systematic deviation from 
the model since we find some of the largest positive and negative 
deviations in the same part of the $(u,v)$ diagram.  We suspect that
the high value of $\chi^2_\nu$ reflects an incomplete accounting 
of the full error budget. 
 
\begin{figure}
\begin{center}
{\includegraphics[angle=90,height=10cm]{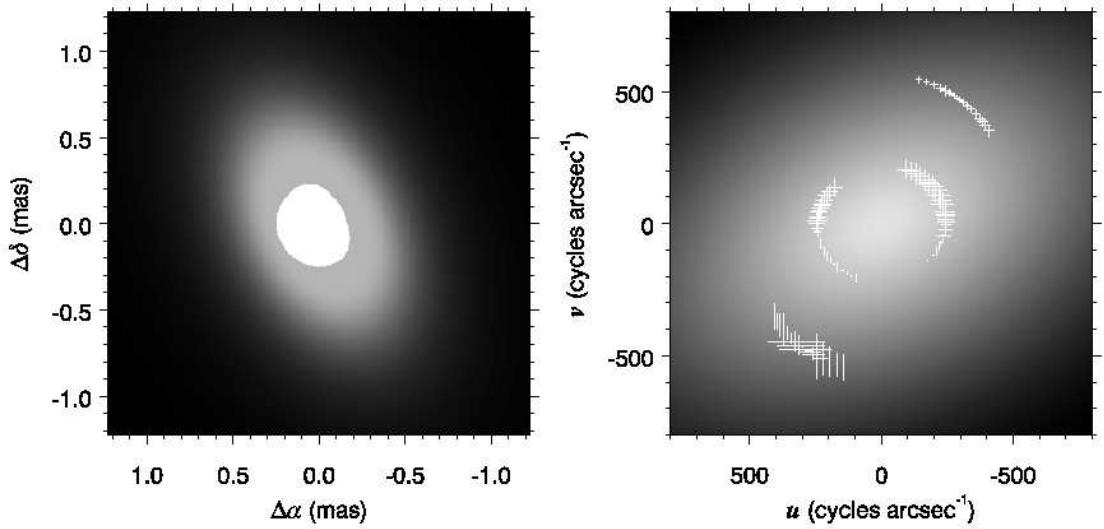}}
\end{center}
\caption{The disk model for $\gamma$~Cas.  The left panel shows
the appearance of the model star and disk in the sky (with the 
spin axis rotated east from north by angle $\alpha$). 
The right panel shows the associated Fourier transform of the
spatial image, which is directly related to the predicted 
interferometric visibility.  The plus signs in the 
upper half of the panel are scaled representations of the 
observed visibility squared at the $(u,v)$ coordinate of observation, 
while the mirror symmetric, thick line segments in the lower half show the 
scaled residuals from the fit (vertical for a positive residual 
and horizontal for a negative residual).  
}
\label{fig6}
\end{figure}

\placefigure{fig6}     

\citet{har00} discovered that $\gamma$~Cas is a single-lined spectroscopic 
binary, and their orbit was confirmed in subsequent Doppler shift measurements
by \citet{mir02}.  We made a fit of the visibilities by including a possible 
hot subdwarf companion that was placed in the model images using the 
spectroscopic ephemeris from \citet{mir02} (Table~1).  This modification made 
a slight improvement in the residuals of the fit, and models with a 
secondary hotter than $T_{\rm eff} = 30$~kK showed somewhat better agreement. 

The profiles of the H$\alpha$, H$\gamma$, and Br$\gamma$ emission lines
are plotted as thick lines in Figure~7.  We also plot the synthetic 
model profiles as thin lines for each feature.  The model profiles of the 
weaker H$\gamma$ and Br$\gamma$ features make a reasonable match of the
observed profiles, but, as noted above, the H$\alpha$ profile is much 
broader and smoother than predicted.  \citet*{rob02} suggest that there 
may be significant turbulence in the inner disk of $\gamma$~Cas that is 
associated with magnetic dynamo processes, and such turbulent motions 
could help to explain the wings observed in H$\alpha$. 

\begin{figure}
\epsscale{0.9}
\plotone{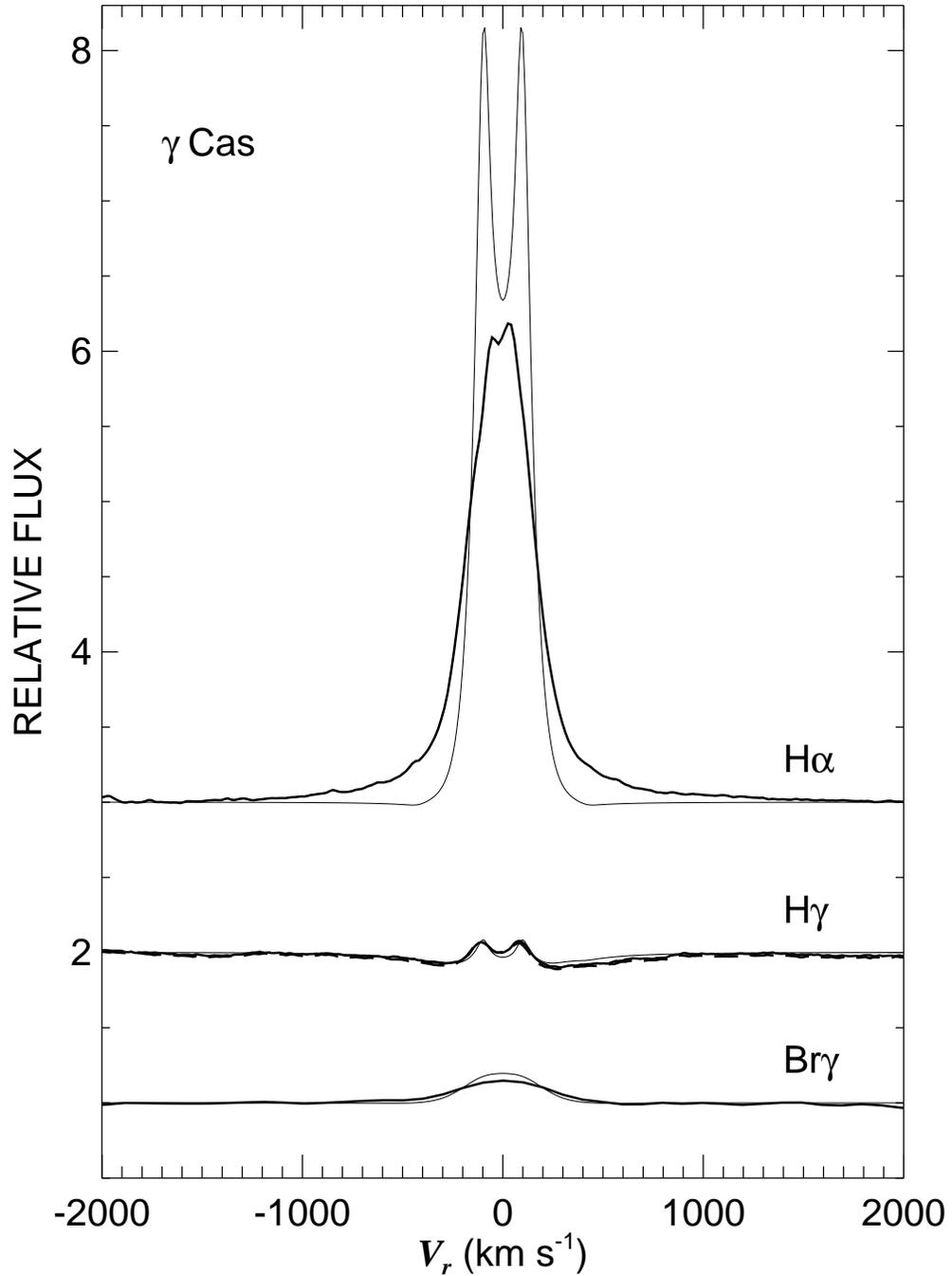}
\caption{Hydrogen line profiles in the spectrum of $\gamma$ Cas 
plotted as a function of heliocentric radial velocity. 
The solid, thick lines indicate spectra made in 2004 while 
the dashed, thick lines correspond to 2005 observations. 
The thin lines represent the model synthetic spectra. 
The continuum levels for H$\alpha$ ({\it top}) and 
H$\gamma$ ({\it middle}) are offset in rectified flux 
by $+2$ and $+1$, respectively, for clarity of presentation.}
\label{fig7}
\end{figure}

\placefigure{fig7}     

\subsection{$\phi$ Per}

The binary system $\phi$~Per is the only one of the four targets
where we known the nature of the secondary and can reliably 
predict its flux contribution in the near-IR \citep{gie98}. 
The companion is a hot subdwarf whose spectrum appears 
clearly in UV spectroscopy from {\it HST}, and from the relative
strengths of its spectral lines, we can estimate with confidence 
its temperature, radius, and hence its near-IR flux contribution. 
Indeed the fit of the CHARA Array visibilities is better for 
the binary model than the single star model (Table~6). 
Here the inclusion of the binary flux leads to a downwards 
revision of the disk size.  We show in left hand panel of Figure~8 
the image of the entire binary for the observation made 
on HJD 2,453,656.9 (orbital phase 0.644), which is similar to the
binary orientation for most of the observations (Table~4). 
The right hand panel shows the visibility image that is now 
modulated with the interference pattern introduced by the 
binary companion.  Since the presence of the secondary does 
affect the visibility pattern, additional interferometric data
should lead to an astrometric orbit that will allow us to test 
whether or not the disk and orbit are co-planar \citep{cla98}. 

\begin{figure}
\begin{center}
{\includegraphics[angle=90,height=10cm]{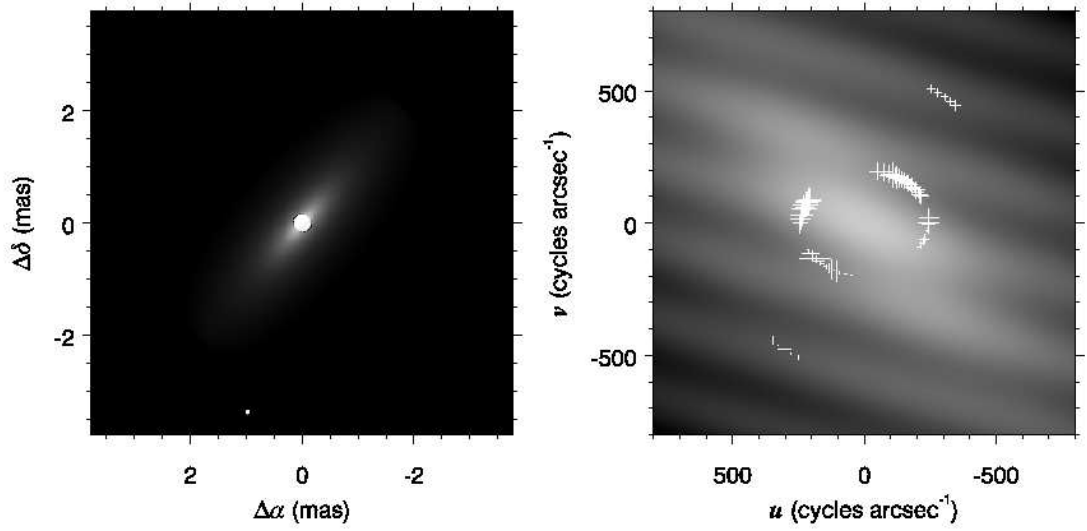}}
\end{center}
\caption{The model image and visibilities for the 
binary star fit of the interferometric data of $\phi$~Per 
in the same format as Fig.~6.  The secondary appears as 
the bright dot in the lower part of the spatial image.}
\label{fig8}
\end{figure}

\placefigure{fig8}     

The position angle of the disk that we find is about $20^\circ$ 
different than that from the recent narrow-band H$\alpha$ measurements 
from \citet{tyc06} (Table~7).  Our solution is somewhat influenced by our 
assumptions about the properties of the secondary's orbit, and 
additional observations at other binary phases will be required 
to settle the issue of the disk and orbit orientations. 
We also find a much lower density exponent $n=1.8$ than 
the value of $n=3.0$ found by \citet{wat87} based on fitting 
the IR-excess from {\it IRAS} data, but their result is sensitive 
to the adopted disk radius.  For example, \citet{wat86} demonstrated 
that a density exponent of $n=2.4$ also fit the IR-excess provided 
the disk radius was limited to $R_d=6.5 R_\odot$, so the difference 
in our results probably hinges on differences in the model assumptions. 

Figure~9 shows the H$\alpha$, H$\gamma$, and Br$\gamma$ profiles for 
$\phi$~Per.  The peaks in the H$\gamma$ profile reversed in strength 
between our observations in 2004 and 2005.  This is an important reminder 
that the disk is time variable and is also probably somewhat asymmetrical
in appearance \citep{hum95,hum01}, properties that are beyond the scope of 
our simple disk models.   Both the H$\alpha$ and Br$\gamma$ line are 
observed to be broader than predicted by the disk model. 

\begin{figure}
\epsscale{0.9}
\plotone{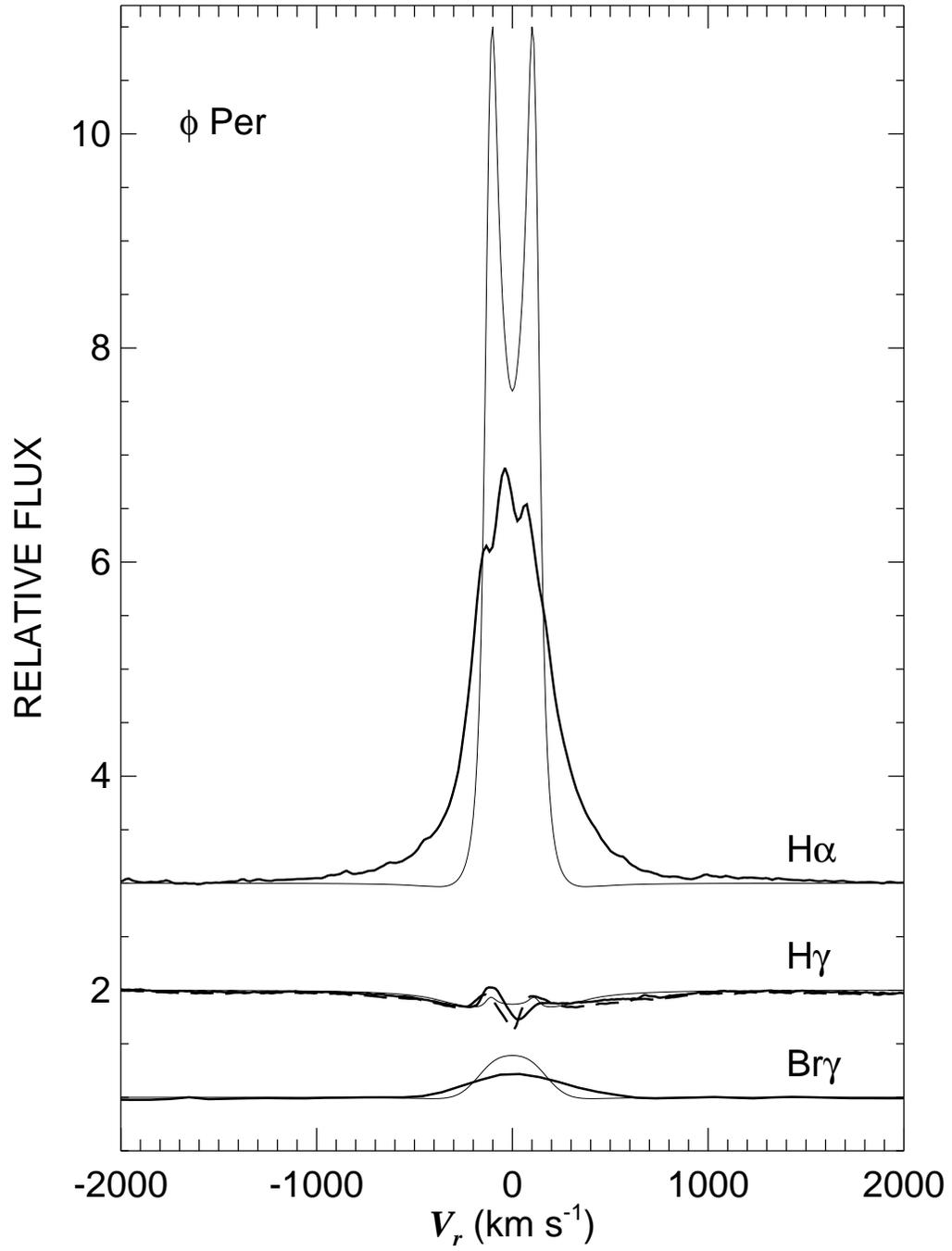}
\caption{Hydrogen line profiles in the spectrum of $\phi$ Per 
plotted in the same format as Fig.~7.}
\label{fig9}
\end{figure}

\placefigure{fig9}     

\subsection{$\zeta$ Tau}

\citet{har84} derived the best radial velocity curve to date for this
single-lined binary, and we adopt his period here but we took the 
epoch from the the more recent spectroscopic study by \citet{kay97}. 
A comparison of the single star and binary fits to the interferometry 
shows essentially no differences in the resulting $\chi^2_\nu$ values, 
so we will discuss the single star results here.  The spatial and 
visibility images resulting from the fit are shown in Figure~10. 
Our results indicate that we are viewing this system close to 
edge-on, and the decline in visibility is especially marked along
the direction of the projected major axis of the disk.  Our 
derived orientation in the sky is in reasonable agreement with 
the H$\alpha$ results from \citet{tyc04}, and the density law agrees 
within errors with that obtained from fitting the IR-excess 
flux \citep{wat87}.  

\begin{figure}
\begin{center}
{\includegraphics[angle=90,height=10cm]{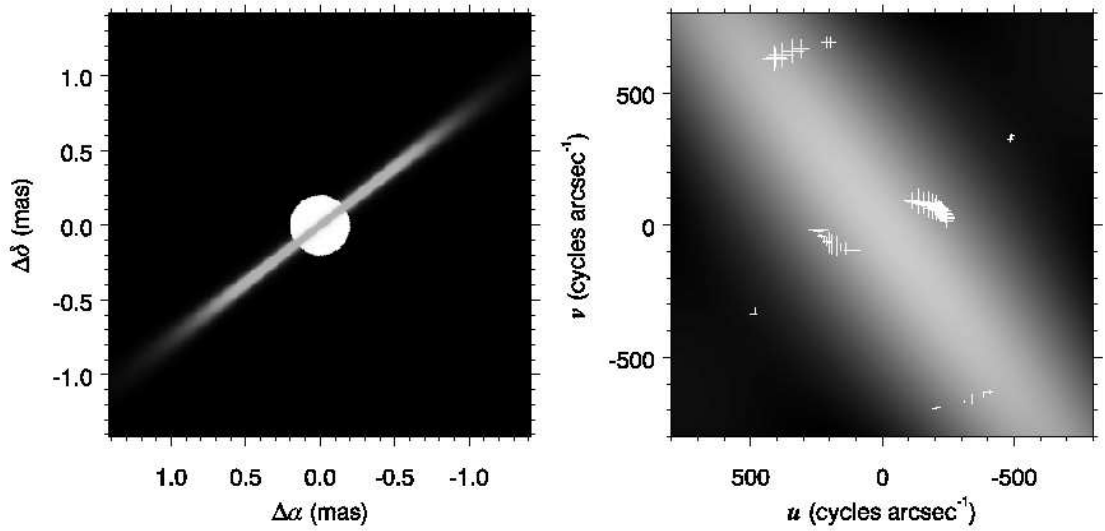}}
\end{center}
\caption{The model image and visibilities for the 
single star fit of the interferometric data of $\zeta$~Tau 
in the same format as Fig.~6.}
\label{fig10}
\end{figure}

\placefigure{fig10}     

The spectral line profiles of $\zeta$~Tau are illustrated in Figure~11.
Again we find that the observed H$\alpha$ line is much broader than 
the model profile, and here we see evidence of a significant 
asymmetry in the disk in the differing heights of the blue and 
red emission peaks.  The long term variations in these asymmetries 
are documented by \citet{guo95}.  \citet{vak98} made H$\alpha$ 
interferometric observations in narrow bands matching the blue and 
red peaks, and their results support the idea that the disk density 
is modulated by a prograde one-armed oscillation. 
The H$\gamma$ line shows a deep central ``shell'' feature that 
is consistent with the formation of such shell lines in Be stars 
observed near $i=90^\circ$ \citep{han96}.  The fact that our model
H$\gamma$ line depth is shallower than the observed depth probably indicates 
that the outer parts of the disk are cooler than assumed in 
our isothermal model.   

\begin{figure}
\epsscale{0.9}
\plotone{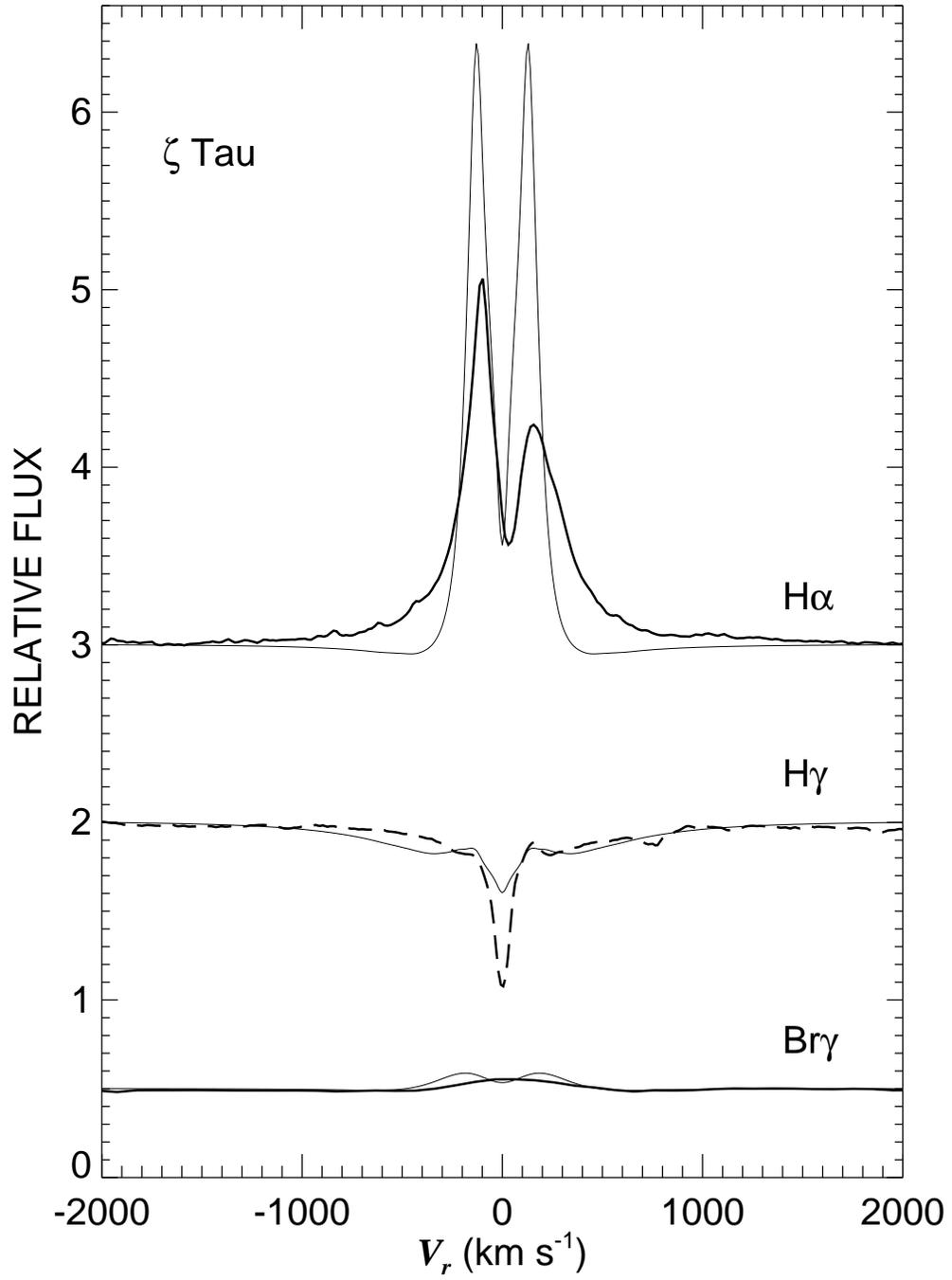}
\caption{Hydrogen line profiles in the spectrum of $\zeta$ Tau 
plotted in the same format as Fig.~7, except that the Br$\gamma$
profile is offset by $-0.5$ in flux for clarity.}
\label{fig11}
\end{figure}

\placefigure{fig11}     

\subsection{$\kappa$ Dra}

\citet{juz91} and \citet{saa05} show that $\kappa$~Dra is 
a low amplitude, single-lined spectroscopic binary with a 
period of 65.6~d.  \citet{saa04} describe a long-term 
cyclic variation of about 22~y that is present in the 
emission lines, and they suggest that this period may 
correspond to the disk's modulation due to an interaction
with a third star in an eccentric orbit.  The semimajor 
axis of the third star would be on the order of 100~mas 
for the adopted distance, but we find no obvious evidence 
(separated fringe packets) for such a companion even at the 
shortest baselines.  Thus, we have confined our test fits
here to the single star and binary star ($P=65$~d) cases. 

The fit for the single star case is odd in several respects. 
The density exponent is unusually small and the inclination 
of $i=90^\circ$ is at odds with estimates from spectroscopy 
($i=35^\circ - 45^\circ$; \citealt{juz91}).  
The binary fit has similar residuals, but it brings the 
density exponent and inclination into better accord 
with expectations.  Thus, we will adopt the binary model 
that is based on a hot, subdwarf companion, but we caution 
that additional interferometric observations are required 
to make reliable estimates for the companion's orbit and
flux contribution. 

The model spatial and visibility images are shown in 
Figure~12, and we see in this case that the binary modulation 
is equal to or greater than the disk variation in importance. 
The CHARA Array measurements presented here are the first 
interferometric observations of this target, so there 
are no H$\alpha$ results for comparison. 
\citet{qui97} found that the intrinsic continuum polarization 
angle is generally perpendicular to the projected disk 
major axis in the sky.  \citet{cla90} 
estimated the intrinsic polarization angle for $\kappa$~Dra as 
$23^\circ$ east from north (in agreement with the estimate of
$27^\circ$ from a recent polarization analysis by K.\
Bjorkman, private communication), and this agrees within 
errors with the position angle of $\alpha = 21^\circ \pm 3^\circ$
from the fit of the interferometric visibilities.  

\begin{figure}
\begin{center}
{\includegraphics[angle=90,height=10cm]{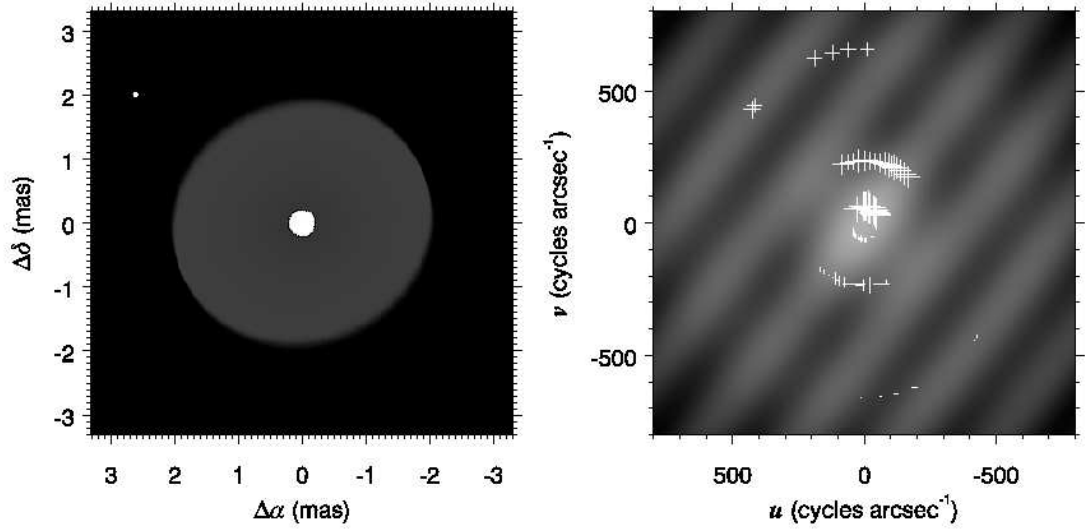}}
\end{center}
\caption{The model image and visibilities for the 
binary star fit of the interferometric data of $\kappa$~Dra 
in the same format as Fig.~6. The secondary appears as 
the bright dot in the upper left part of the spatial image.}
\label{fig12}
\end{figure}

\placefigure{fig12}     

The line profiles are shown in Figure~13.  We find that the 
H$\alpha$ emission strength is approximately that expected 
for the epoch of our observations and the long term variation 
(see Fig.~4 of \citealt{saa04}).  Once again the H$\alpha$ 
profile (and perhaps the H$\gamma$ profile) appear to be broader 
than predicted by the Keplerian model.  

\begin{figure}
\epsscale{0.9}
\plotone{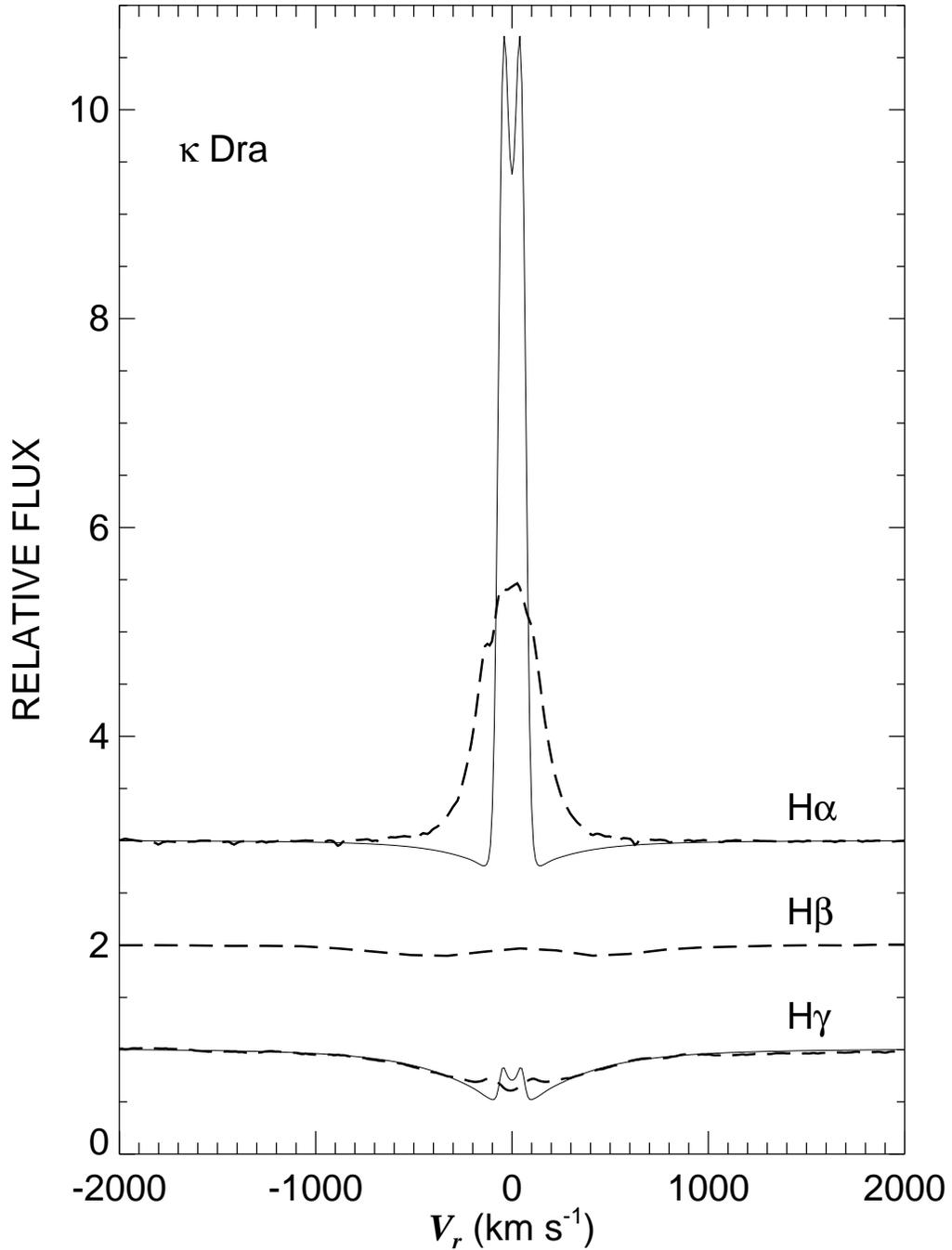}
\caption{Hydrogen line profiles in the spectrum of $\kappa$ Dra 
plotted in the same format as Fig.~7.  We lack observations 
of Br$\gamma$ for this star but we include our low resolution 
spectrum of H$\beta$ instead (without a synthetic comparison).   
The continuum levels for H$\alpha$ ({\it top}) and 
H$\beta$ ({\it middle}) are offset in rectified flux 
by $+2$ and $+1$, respectively, for clarity of presentation.}
\label{fig13}
\end{figure}

\placefigure{fig13}     


\section{Conclusions}                               

The CHARA Array observations represent the first $K^\prime$ 
interferometric measurements of northern Be stars, and we find 
that the disks of all four targets are at least partially 
resolved.  The derived disk inclination angle and position 
angle in the sky generally agree with similar estimates 
from narrow-band H$\alpha$ interferometry where available.  
We find, for example, that the disk of $\zeta$~Tau appears 
with a nearly edge-on orientation, which is consistent with the 
prediction for Be stars with shell spectral features \citep{han96}. 
All four of the targets are close binaries with faint companions, 
and these companions appear to influence the $K^\prime$ visibilities 
of $\phi$~Per and $\kappa$~Dra, an interesting but complicating 
factor for the interpretation of the disk properties. 
We find, for example, that the radial gradient in disk density
(density exponent $n$ in Table~6) 
is smaller in binaries with smaller semimajor axes, which 
suggests that binary companions do influence the disk properties.  

We developed a simple Keplerian disk model to create synthetic 
angular images of the disks, and we used these models to 
create model interferometric visibility patterns and 
spectral emission line profiles for comparison with our
observations.   These models assume isothermal disks 
and ignore scattering in the radiative transfer process, 
two significant limitations that should be addressed in 
a more complete model.  The disks are characterized by 
two physical parameters, the base density and density exponent, 
and two observational parameters, the disk normal inclination
and position angle.  We fit the CHARA Array visibilities for 
all four parameters for both single and binary star models, 
and our results are in good agreement with prior studies of
the IR-excess in these stars.  The parameters derived from 
the interferometry lead to model emission lines that match 
the observed ones if the base densities are reduced by a factor 
of about 1.7~dex.  This mismatch is not unexpected considering 
the many simplifying assumptions made in the model. 

We find that the angular diameters of the Be disks in the $K^\prime$-band 
are consistently smaller than those found from H$\alpha$ interferometry. 
We think that this difference results from the larger H$\alpha$ 
opacity and the hydrogen ionization structure in the disk.  
The $K^\prime$ disk emission comes primarily from free-free processes in the 
mainly ionized gas, and the free-free optical depth will vary with disk 
radius according to the square of the density.  In the 
thin disk approximation, the optical depth at radial distance $R$
will vary as $\tau \propto \rho_0^2 ~R^{-2n+3/2} / \cos i$
(see eq.\ [2] and [3]).  On the other hand, the H$\alpha$ emission 
depends on the relatively low neutral hydrogen number density,
and we assumed above that this varies inversely with the stellar 
dilution factor (eq.\ [9]; $W(r)\propto R^{-2}$ at large radius and $Z=0$).
Thus, we expect the H$\alpha$ optical depth to vary 
approximately as $\tau \propto \rho_0^2 ~R^{-2n+3/2+2} / \cos i$, 
i.e., with a shallower effective density exponent of $n-1$.
Consequently, the H$\alpha$ optical depth reduction with 
increasing disk radius will be less pronounced than in 
the case of the IR optical depth, resulting in a 
larger spatial extension in the H$\alpha$ emission
compared to that for the $K^\prime$ emission. 

This expectation is borne out in the predicted spatial images. 
We show an example of the derived spatial structure in Figure~14 
for our single star model of $\gamma$~Cas.  The solid line 
shows the summed $K^\prime$-band intensity of the image 
projected onto the major axis of the disk and 
measured in stellar radii from the center of the star.  
We used our model to create an 
H$\alpha$ image adopting the base density from the fit 
of the observed emission line (Table~8).  The intensity 
was formed by summing across a 2.8~nm wavelength band centered 
on H$\alpha$ in order to match the H$\alpha$ interferometric
observations of \citet{tyc06}.  
Note that changing the adopted bandwidth will simply  
result in a rescaling of the contrast between the central star
and the surrounding disk emission.  The disk image appears 
darker along the line of nodes and along the line of sight because
of the minimal shear broadening that occurs at those disk 
azimuthal angles and the subsequent reduction in the range of 
wavelength over which H$\alpha$ is optically thick.  The dashed line in Figure~14
shows the summed H$\alpha$ image intensity projected onto the major axis. 
We see that H$\alpha$ emission is indeed more extended along the major axis 
and that the half maximum intensity radii are located at $2.6 R_\star$ and 
$7.4 R_\star$ for the $K^\prime$ band and H$\alpha$ summed intensities, 
respectively.  The ratio of these half maximum intensity radii is 
0.35, which is comparable to the observed ratio of 0.55 of the 
FWHM of the major axes of the $K^\prime$-band 
and H$\alpha$ emission in the Gaussian elliptical fits 
(Table~7).  Thus, the differences in the radial exponent dependence of
density squared and the neutral hydrogen fraction may partially explain 
the larger appearance of the H$\alpha$ disks. 

\begin{figure}
\epsscale{1.}
\begin{center}
{\includegraphics[angle=90,height=10cm]{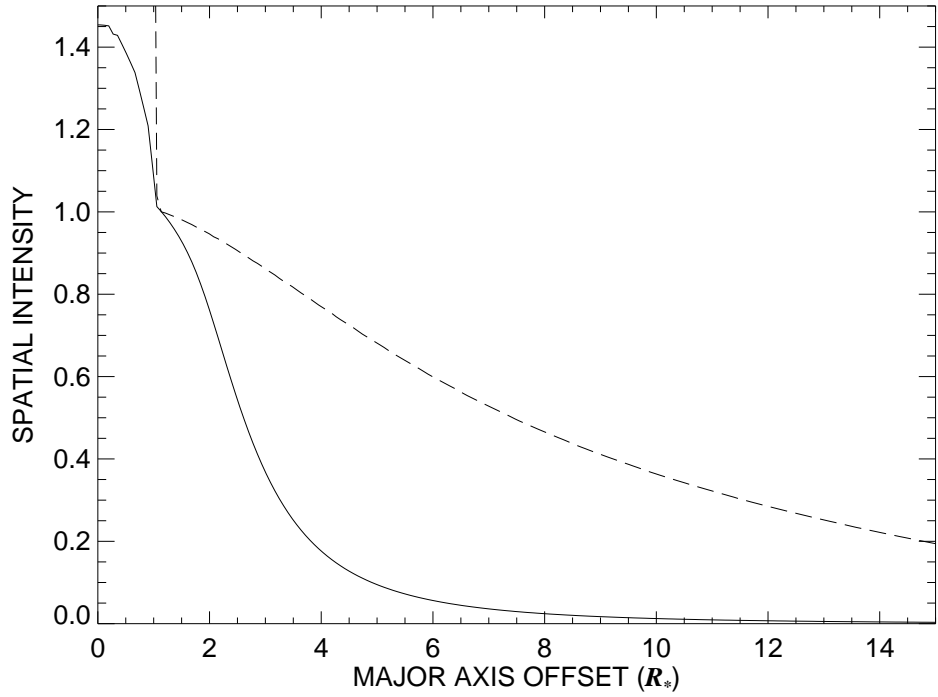}}
\end{center}
\caption{The summed spatial intensities projected onto the 
major axis for our single star model of $\gamma$~Cas. The solid 
line shows the $K^\prime$-band summed intensity while the dashed line
shows the summed intensity for an H$\alpha$ image integrated over a 
2.8~nm bandwidth.  Both are renormalized for convenience to 
their respective summed intensity at $R = 1.14 R_\star$.}
\label{fig14}
\end{figure}

\placefigure{fig14}     

The CHARA Array observations demonstrate the promise of these 
kinds of measurements for our understanding of Be stars.  
Many Be stars were probably spun up by mass transfer in a 
past close interaction, and future observations over a range
in $(u,v)$ and orbital phase coverage will reveal the nature of
their companions.   The disks of Be stars are driven by dynamic
processes and evolve over timescales of months to years, so 
long-term observations will help elucidate the physics of 
disk formation and dispersal.  There is ample evidence that 
the disks are structured and prone to the formation of spiral 
arm features \citep{oka02}.  Furthermore, in Be binaries like 
$\phi$~Per, the portion of the disk closest to the hot companion
may appear brighter due to local heating by the companion's 
radiation \citep{gie98,hum01}.  The techniques of phase closure 
using multiple telescopes in long baseline interferometry \citep{kra05}
will soon give us the means to study these disk processes directly 
among the nearby Be stars.


\acknowledgments

We are grateful for helpful advice received from John Monnier, 
Christopher Tycner, Philippe Stee, and Andrew Boden.  
This material is based on work supported by the National Science
Foundation under grants AST--0205297, AST--0307562, and AST--0401460.
We are grateful for travel support (for M.\ V.\ McSwain) 
from Hokkai-Gakuen University in Sapporo where 
preliminary results from this study were presented. 
Institutional support has been provided from the GSU College
of Arts and Sciences and from the Research Program Enhancement
fund of the Board of Regents of the University System of Georgia,
administered through the GSU Office of the Vice President for Research.
We gratefully acknowledge all this support.





\clearpage


\begin{deluxetable}{lcccc}
\tablewidth{0pc}
\tablecaption{Adopted Stellar Parameters\label{tab1}}
\tablehead{
\colhead{Parameter}    &
\colhead{$\gamma$ Cas} &
\colhead{$\phi$ Per}   &
\colhead{$\zeta$ Tau}  &
\colhead{$\kappa$ Dra} 
}
\startdata
Spectral Classification \dotfill & B0.5 Ve  & B0.5 III-Ve & B2 IIIpe & B5-6 IIIe \\
$\pi$ (mas)             \dotfill &  5.32    &  4.55    &  7.82    &  6.55    \\
$R_1$ ($R_\odot$)       \dotfill &  10.0    &   7.0    &   5.5    &   6.4    \\
$M_1$ ($M_\odot$)       \dotfill &  15.5    &   9.3    &  11.2    &   4.8    \\
$T_{1~\rm eff}$ (K)     \dotfill & 28840    & 29300    & 19000    & 14000    \\
$V\sin i$ (km s$^{-1}$) \dotfill &  295     &  410     &  320     &  170     \\
$R_2$ ($R_\odot$)\tablenotemark{a}       \dotfill & 1.0      & 1.33     & 1.0      & 1.0      \\
$M_2$ ($M_\odot$)\tablenotemark{a}       \dotfill & 1.0      & 1.1      & 1.0      & 0.8      \\
$T_{2~\rm eff}$ (K)\tablenotemark{a}     \dotfill & 30000    & 53000    & 30000    & 30000    \\
$P$ (d)                 \dotfill & 203.59   & 126.67   & 132.91   &  61.56   \\
$T_{S+}$ (HJD--2,400,000)\dotfill& 50654.3  & 50155.1  & 46417.3  & 50011.0  \\
$a_1 \sin i$ ($R_\odot$)\dotfill & 15.279   & 24.942   & 27.562   & 10.078   \\
$R_d$ ($R_\odot$)       \dotfill &  214     &  130     &  146     &   67     \\
References              \dotfill & 1, 2     & 3        & 4, 5, 6  &  7       \\
\enddata
\tablenotetext{a}{Example parameters for an assumed subdwarf companion for
all but $\phi$ Per where the estimates are from \citet{gie98}.}
\tablerefs{
1. \citet{har00};
2. \citet{mir02};
3. \citet{gie98};
4. \citet{har84};
5. \citet{yan90};
6. \citet{kay97};
7. \citet{saa05}.
}
\end{deluxetable}

\clearpage


\begin{deluxetable}{lcccc}
\tablewidth{0pc}
\tablecaption{CHARA Array Observations\label{tab2}}
\tablehead{
\colhead{Target} &
\colhead{Calibrator} &
\colhead{Baseline} &
\colhead{Date} &
\colhead{No.\ of} \\
\colhead{Name} &
\colhead{Name} &
\colhead{(m)} &
\colhead{(UT)} &
\colhead{Sets}
}
\startdata
$\gamma$ Cas  & HD 6210   & W1/S2 (249)    & 2003 Oct 07 & \phn5 \\
$\gamma$ Cas  & HD 6210   & W1/S2 (249)    & 2003 Oct 10 & \phn6 \\
$\gamma$ Cas  & HD 6210   & W1/S2 (249)    & 2003 Oct 14 & \phn5 \\
$\gamma$ Cas  & HD 6210   & W1/S2 (249)    & 2003 Oct 16 & \phn6 \\
$\gamma$ Cas  & HD 6210   & W1/W2 (111)    & 2005 Oct 12 &    12 \\
$\gamma$ Cas  & HD 6210   & W1/W2 (111)    & 2005 Oct 13 &    11 \\
$\gamma$ Cas  & HD 6210   & W1/W2 (111)    & 2005 Oct 15 &    11 \\
$\phi$ Per    & HD 6961   & W1/S2 (249)    & 2003 Oct 03 & \phn5 \\
$\phi$ Per    & HD 6961   & W1/W2 (111)    & 2003 Oct 12 &    10 \\
$\phi$ Per    & HD 6961   & W1/W2 (111)    & 2003 Oct 13 & \phn7 \\
$\phi$ Per    & HD 6961   & W1/W2 (111)    & 2003 Oct 14 &    14 \\
$\zeta$ Tau   & HD 43042  & S1/E1 (331)    & 2004 Dec 21 & \phn6 \\
$\zeta$ Tau   & HD 32977  & S1/E1 (331)    & 2004 Dec 22 & \phn1 \\
$\zeta$ Tau   & HD 43042  & W1/S1 (279)    & 2005 Apr 01 & \phn2 \\
$\zeta$ Tau   & HD 43042  & W1/W2 (111)    & 2005 Dec 04 & \phn8 \\
$\zeta$ Tau   & HD 43042  & W1/W2 (111)    & 2005 Dec 05 & \phn9 \\
$\kappa$ Dra  & HD 107193 & S1/S2 (34)\phn & 2005 Apr 02 &    13 \\
$\kappa$ Dra  & HD 107193 & S1/E1 (331)    & 2005 Apr 06 & \phn5 \\
$\kappa$ Dra  & HD 107193 & S1/E1 (331)    & 2005 Apr 13 & \phn1 \\
$\kappa$ Dra  & HD 107193 & W1/W2 (111)    & 2005 Dec 04 & \phn3 \\
$\kappa$ Dra  & HD 107193 & W1/W2 (111)    & 2005 Dec 05 &    13 \\
\enddata
\end{deluxetable}

\clearpage


\begin{deluxetable}{lcccccc}
\tablewidth{0pc}
\tablecaption{Calibrator Star Angular Diameters\label{tab3}}
\tablehead{
\colhead{Calibrator} &
\colhead{$T_{\rm eff}$} &
\colhead{$\log g$} &
\colhead{Spectral} &
\colhead{$E(B-V)$} &
\colhead{$\theta_{LD}$} &
\colhead{} \\
\colhead{Name} &
\colhead{(K)} &
\colhead{(cm s$^{-2}$)} &
\colhead{Classification} &
\colhead{(mag)} &
\colhead{(mas)} &
\colhead{Ref.}
}
\startdata
HD 6210   &  6087 &  3.73 & F6 V     & 0.00 &  $0.523 \pm 0.027$ & 1, 2, 3, 4 \\
HD 6961   &  7762 &  3.80 & A7 V var & 0.00 &  $0.608 \pm 0.019$ & 2 \\
HD 32977  &  8511 &  4.19 & A5 V     & 0.00 &  $0.344 \pm 0.033$ & 2 \\
HD 43042  &  6556 &  4.28 & F6 V     & 0.00 &  $0.590 \pm 0.016$ & 3 \\
HD 107193 &  8710 &  3.93 & A0 Vn    & 0.04 &  $0.308 \pm 0.005$ & 2 \\
\enddata
\tablerefs{
1. \citet{cay97};
2. \citet{all99};
3. \citet{lam04};
4. \citet{nor04}.
}
\end{deluxetable}
\clearpage


\begin{deluxetable}{lcccccc}
\tablewidth{0pc}
\tablecaption{Calibrated $K^\prime$ Visibilities\label{tab4}}
\tablehead{
\colhead{Target} &
\colhead{Date} &
\colhead{Orbital} &
\colhead{Telescope} &
\colhead{Baseline} &
\colhead{Pos. Ang.} &
\colhead{} \\
\colhead{Name} &
\colhead{(HJD-2,400,000)} &
\colhead{Phase} &
\colhead{Pair} &
\colhead{(m)} &
\colhead{(deg)} &
\colhead{Visibility\tablenotemark{a}}
}
\startdata
$\gamma$ Cas & 52919.721  & 0.127 & W1/S2 &  247.37 &  345.28 & $0.445\pm0.016$ \\
$\gamma$ Cas & 52919.731  & 0.127 & W1/S2 &  247.09 &  342.34 & $0.448\pm0.016$ \\
$\gamma$ Cas & 52919.742  & 0.127 & W1/S2 &  246.71 &  339.22 & $0.490\pm0.017$ \\
$\gamma$ Cas & 52919.751  & 0.128 & W1/S2 &  246.28 &  336.45 & $0.516\pm0.017$ \\
$\gamma$ Cas & 52919.759  & 0.128 & W1/S2 &  245.87 &  334.14 & $0.608\pm0.019$ \\
$\gamma$ Cas & 52922.794  & 0.142 & W1/S2 &  242.31 &  321.83 & $0.496\pm0.016$ \\
$\gamma$ Cas & 52922.804  & 0.143 & W1/S2 &  241.12 &  319.06 & $0.522\pm0.016$ \\
$\gamma$ Cas & 52922.812  & 0.143 & W1/S2 &  240.06 &  316.88 & $0.595\pm0.015$ \\
$\gamma$ Cas & 52922.819  & 0.143 & W1/S2 &  239.06 &  315.01 & $0.593\pm0.018$ \\
$\gamma$ Cas & 52922.825  & 0.143 & W1/S2 &  238.11 &  313.36 & $0.592\pm0.024$ \\
$\gamma$ Cas & 52922.834  & 0.143 & W1/S2 &  236.47 &  310.77 & $0.624\pm0.018$ \\
$\gamma$ Cas & 52926.747  & 0.162 & W1/S2 &  245.49 &  332.32 & $0.458\pm0.019$ \\
$\gamma$ Cas & 52926.754  & 0.162 & W1/S2 &  244.98 &  330.14 & $0.381\pm0.014$ \\
$\gamma$ Cas & 52926.763  & 0.162 & W1/S2 &  244.28 &  327.54 & $0.424\pm0.018$ \\
$\gamma$ Cas & 52926.770  & 0.162 & W1/S2 &  243.70 &  325.68 & $0.509\pm0.016$ \\
$\gamma$ Cas & 52926.777  & 0.162 & W1/S2 &  243.05 &  323.78 & $0.533\pm0.018$ \\
$\gamma$ Cas & 52928.731  & 0.172 & W1/S2 &  246.08 &  335.28 & $0.281\pm0.026$ \\
$\gamma$ Cas & 52928.740  & 0.172 & W1/S2 &  245.57 &  332.66 & $0.266\pm0.023$ \\
$\gamma$ Cas & 52928.746  & 0.172 & W1/S2 &  245.20 &  331.04 & $0.313\pm0.016$ \\
$\gamma$ Cas & 52928.753  & 0.172 & W1/S2 &  244.64 &  328.83 & $0.287\pm0.010$ \\
$\gamma$ Cas & 52928.761  & 0.172 & W1/S2 &  244.00 &  326.62 & $0.298\pm0.011$ \\
$\gamma$ Cas & 52928.770  & 0.172 & W1/S2 &  243.21 &  324.23 & $0.303\pm0.009$ \\
$\gamma$ Cas & 53655.813  & 0.743 & W1/W2 &  107.91 &  100.65 & $0.826\pm0.019$ \\
$\gamma$ Cas & 53655.826  & 0.743 & W1/W2 &  107.91 &\phn96.49& $0.790\pm0.018$ \\
$\gamma$ Cas & 53655.842  & 0.743 & W1/W2 &  107.70 &\phn91.62& $0.791\pm0.021$ \\
$\gamma$ Cas & 53655.854  & 0.743 & W1/W2 &  107.34 &\phn87.65& $0.816\pm0.019$ \\
$\gamma$ Cas & 53655.866  & 0.743 & W1/W2 &  106.84 &\phn83.78& $0.807\pm0.019$ \\
$\gamma$ Cas & 53655.879  & 0.743 & W1/W2 &  106.19 &\phn79.89& $0.756\pm0.019$ \\
$\gamma$ Cas & 53655.891  & 0.743 & W1/W2 &  105.37 &\phn75.74& $0.781\pm0.017$ \\
$\gamma$ Cas & 53655.904  & 0.743 & W1/W2 &  104.45 &\phn71.72& $0.777\pm0.017$ \\
$\gamma$ Cas & 53655.916  & 0.743 & W1/W2 &  103.41 &\phn67.63& $0.779\pm0.018$ \\
$\gamma$ Cas & 53655.929  & 0.744 & W1/W2 &  102.19 &\phn63.12& $0.782\pm0.019$ \\
$\gamma$ Cas & 53655.942  & 0.744 & W1/W2 &  100.93 &\phn58.68& $0.735\pm0.018$ \\
$\gamma$ Cas & 53655.961  & 0.744 & W1/W2 &\phn99.00&\phn52.04& $0.741\pm0.021$ \\
$\gamma$ Cas & 53656.652  & 0.747 & W1/W2 &\phn99.51&  155.13 & $0.839\pm0.024$ \\
$\gamma$ Cas & 53656.664  & 0.747 & W1/W2 &  100.24 &  150.39 & $0.809\pm0.023$ \\
$\gamma$ Cas & 53656.678  & 0.747 & W1/W2 &  101.09 &  145.40 & $0.830\pm0.029$ \\
$\gamma$ Cas & 53656.691  & 0.747 & W1/W2 &  101.97 &  140.64 & $0.792\pm0.027$ \\
$\gamma$ Cas & 53656.707  & 0.747 & W1/W2 &  103.12 &  134.75 & $0.829\pm0.032$ \\
$\gamma$ Cas & 53656.720  & 0.747 & W1/W2 &  104.04 &  130.07 & $0.833\pm0.024$ \\
$\gamma$ Cas & 53656.734  & 0.747 & W1/W2 &  104.90 &  125.58 & $0.859\pm0.021$ \\
$\gamma$ Cas & 53656.747  & 0.748 & W1/W2 &  105.70 &  121.17 & $0.833\pm0.021$ \\
$\gamma$ Cas & 53656.762  & 0.748 & W1/W2 &  106.52 &  116.04 & $0.878\pm0.022$ \\
$\gamma$ Cas & 53656.775  & 0.748 & W1/W2 &  107.10 &  111.71 & $0.840\pm0.020$ \\
$\gamma$ Cas & 53656.788  & 0.748 & W1/W2 &  107.51 &  107.71 & $0.884\pm0.025$ \\
$\gamma$ Cas & 53658.701  & 0.757 & W1/W2 &  103.11 &  134.83 & $0.875\pm0.021$ \\
$\gamma$ Cas & 53658.714  & 0.757 & W1/W2 &  103.97 &  130.47 & $0.830\pm0.020$ \\
$\gamma$ Cas & 53658.727  & 0.757 & W1/W2 &  104.86 &  125.83 & $0.851\pm0.018$ \\
$\gamma$ Cas & 53658.741  & 0.757 & W1/W2 &  105.67 &  121.35 & $0.878\pm0.019$ \\
$\gamma$ Cas & 53658.814  & 0.758 & W1/W2 &  107.93 &\phn97.77& $0.779\pm0.021$ \\
$\gamma$ Cas & 53658.827  & 0.758 & W1/W2 &  107.82 &\phn93.74& $0.742\pm0.021$ \\
$\gamma$ Cas & 53658.840  & 0.758 & W1/W2 &  107.54 &\phn89.60& $0.740\pm0.023$ \\
$\gamma$ Cas & 53658.854  & 0.758 & W1/W2 &  107.01 &\phn84.98& $0.704\pm0.024$ \\
$\gamma$ Cas & 53658.876  & 0.758 & W1/W2 &  105.88 &\phn78.23& $0.684\pm0.022$ \\
$\gamma$ Cas & 53658.892  & 0.758 & W1/W2 &  104.77 &\phn73.07& $0.740\pm0.022$ \\
$\gamma$ Cas & 53658.906  & 0.758 & W1/W2 &  103.61 &\phn68.38& $0.754\pm0.025$ \\
$\phi$ Per   & 52915.807  & 0.794 & W1/S2 &  249.35 &  333.52 & $0.533\pm0.019$ \\
$\phi$ Per   & 52915.818  & 0.794 & W1/S2 &  249.21 &  330.46 & $0.520\pm0.016$ \\
$\phi$ Per   & 52915.829  & 0.794 & W1/S2 &  248.93 &  327.43 & $0.475\pm0.015$ \\
$\phi$ Per   & 52915.840  & 0.794 & W1/S2 &  248.50 &  324.62 & $0.540\pm0.016$ \\
$\phi$ Per   & 52915.850  & 0.794 & W1/S2 &  247.94 &  322.06 & $0.586\pm0.020$ \\
$\phi$ Per   & 53655.657  & 0.635 & W1/W2 &\phn89.21&  165.54 & $0.771\pm0.028$ \\
$\phi$ Per   & 53655.673  & 0.635 & W1/W2 &\phn90.30&  159.13 & $0.753\pm0.022$ \\
$\phi$ Per   & 53655.687  & 0.635 & W1/W2 &\phn91.70&  153.00 & $0.783\pm0.027$ \\
$\phi$ Per   & 53655.702  & 0.635 & W1/W2 &\phn93.31&  147.18 & $0.743\pm0.020$ \\
$\phi$ Per   & 53655.715  & 0.635 & W1/W2 &\phn94.94&  142.08 & $0.789\pm0.021$ \\
$\phi$ Per   & 53655.728  & 0.635 & W1/W2 &\phn96.63&  137.21 & $0.724\pm0.019$ \\
$\phi$ Per   & 53655.743  & 0.635 & W1/W2 &\phn98.54&  132.02 & $0.749\pm0.020$ \\
$\phi$ Per   & 53655.758  & 0.635 & W1/W2 &  100.43 &  127.03 & $0.753\pm0.018$ \\
$\phi$ Per   & 53655.772  & 0.635 & W1/W2 &  102.16 &  122.38 & $0.761\pm0.019$ \\
$\phi$ Per   & 53655.789  & 0.636 & W1/W2 &  104.03 &  117.05 & $0.724\pm0.018$ \\
$\phi$ Per   & 53656.864  & 0.644 & W1/W2 &  107.89 &\phn94.57& $0.836\pm0.025$ \\
$\phi$ Per   & 53656.878  & 0.644 & W1/W2 &  107.55 &\phn90.58& $0.866\pm0.024$ \\
$\phi$ Per   & 53656.892  & 0.644 & W1/W2 &  106.92 &\phn86.85& $0.840\pm0.024$ \\
$\phi$ Per   & 53656.905  & 0.644 & W1/W2 &  106.00 &\phn83.15& $0.827\pm0.024$ \\
$\phi$ Per   & 53656.932  & 0.645 & W1/W2 &  103.25 &\phn75.29& $0.820\pm0.031$ \\
$\phi$ Per   & 53656.947  & 0.645 & W1/W2 &  101.27 &\phn70.73& $0.875\pm0.026$ \\
$\phi$ Per   & 53656.960  & 0.645 & W1/W2 &\phn99.35&\phn66.67& $0.887\pm0.026$ \\
$\phi$ Per   & 53657.690  & 0.651 & W1/W2 &\phn92.54&  149.83 & $0.937\pm0.039$ \\
$\phi$ Per   & 53657.703  & 0.651 & W1/W2 &\phn94.10&  144.64 & $0.880\pm0.031$ \\
$\phi$ Per   & 53657.717  & 0.651 & W1/W2 &\phn95.89&  139.29 & $0.744\pm0.024$ \\
$\phi$ Per   & 53657.733  & 0.651 & W1/W2 &\phn97.90&  133.73 & $0.684\pm0.022$ \\
$\phi$ Per   & 53657.746  & 0.651 & W1/W2 &\phn99.59&  129.25 & $0.643\pm0.021$ \\
$\phi$ Per   & 53657.759  & 0.651 & W1/W2 &  101.25 &  124.84 & $0.554\pm0.018$ \\
$\phi$ Per   & 53657.775  & 0.651 & W1/W2 &  103.18 &  119.53 & $0.620\pm0.018$ \\
$\phi$ Per   & 53657.790  & 0.651 & W1/W2 &  104.65 &  115.13 & $0.707\pm0.021$ \\
$\phi$ Per   & 53657.879  & 0.652 & W1/W2 &  107.40 &\phn89.50& $0.725\pm0.027$ \\
$\phi$ Per   & 53657.892  & 0.652 & W1/W2 &  106.71 &\phn85.92& $0.826\pm0.020$ \\
$\phi$ Per   & 53657.905  & 0.652 & W1/W2 &  105.77 &\phn82.34& $0.861\pm0.021$ \\
$\phi$ Per   & 53657.925  & 0.652 & W1/W2 &  103.73 &\phn76.48& $0.924\pm0.017$ \\
$\phi$ Per   & 53657.938  & 0.653 & W1/W2 &  102.16 &\phn72.72& $0.931\pm0.016$ \\
$\phi$ Per   & 53657.950  & 0.653 & W1/W2 &  100.36 &\phn68.77& $0.939\pm0.016$ \\
$\zeta$ Tau  & 53360.760  & 0.242 & S1/E1 &  329.84 &\phn33.00& $0.924\pm0.037$ \\
$\zeta$ Tau  & 53360.774  & 0.242 & S1/E1 &  328.55 &\phn30.86& $0.918\pm0.034$ \\
$\zeta$ Tau  & 53360.795  & 0.242 & S1/E1 &  325.81 &\phn27.32& $0.898\pm0.036$ \\
$\zeta$ Tau  & 53360.807  & 0.242 & S1/E1 &  323.78 &\phn24.94& $0.789\pm0.038$ \\
$\zeta$ Tau  & 53360.845  & 0.242 & S1/E1 &  317.39 &\phn17.02& $0.615\pm0.021$ \\
$\zeta$ Tau  & 53360.850  & 0.242 & S1/E1 &  316.65 &\phn15.94& $0.609\pm0.020$ \\
$\zeta$ Tau  & 53361.760  & 0.249 & S1/E1 &  329.62 &\phn32.58& $0.813\pm0.038$ \\
$\zeta$ Tau  & 53461.662  & 0.001 & W1/S1 &  261.82 &  304.56 & $0.368\pm0.019$ \\
$\zeta$ Tau  & 53461.673  & 0.001 & W1/S1 &  256.24 &  303.94 & $0.438\pm0.019$ \\
$\zeta$ Tau  & 53708.795  & 0.860 & W1/W2 &\phn92.98&  108.47 & $0.748\pm0.024$ \\
$\zeta$ Tau  & 53708.809  & 0.860 & W1/W2 &\phn96.78&  106.04 & $0.719\pm0.023$ \\
$\zeta$ Tau  & 53708.822  & 0.860 & W1/W2 &  100.05 &  103.84 & $0.803\pm0.022$ \\
$\zeta$ Tau  & 53708.838  & 0.861 & W1/W2 &  103.33 &  101.35 & $0.775\pm0.028$ \\
$\zeta$ Tau  & 53708.853  & 0.861 & W1/W2 &  105.68 &\phn99.12& $0.746\pm0.022$ \\
$\zeta$ Tau  & 53708.869  & 0.861 & W1/W2 &  107.24 &\phn96.97& $0.767\pm0.022$ \\
$\zeta$ Tau  & 53708.876  & 0.861 & W1/W2 &  107.67 &\phn95.97& $0.716\pm0.021$ \\
$\zeta$ Tau  & 53708.890  & 0.861 & W1/W2 &  107.93 &\phn94.14& $0.675\pm0.020$ \\
$\zeta$ Tau  & 53709.718  & 0.867 & W1/W2 &\phn65.27&  129.49 & $0.742\pm0.032$ \\
$\zeta$ Tau  & 53709.734  & 0.867 & W1/W2 &\phn71.89&  123.24 & $0.924\pm0.029$ \\
$\zeta$ Tau  & 53709.751  & 0.867 & W1/W2 &\phn78.20&  118.27 & $0.868\pm0.028$ \\
$\zeta$ Tau  & 53709.765  & 0.868 & W1/W2 &\phn83.77&  114.39 & $0.946\pm0.029$ \\
$\zeta$ Tau  & 53709.782  & 0.868 & W1/W2 &\phn89.40&  110.74 & $0.927\pm0.028$ \\
$\zeta$ Tau  & 53709.795  & 0.868 & W1/W2 &\phn93.81&  107.95 & $0.948\pm0.032$ \\
$\zeta$ Tau  & 53709.809  & 0.868 & W1/W2 &\phn97.51&  105.57 & $0.815\pm0.021$ \\
$\zeta$ Tau  & 53709.823  & 0.868 & W1/W2 &  100.87 &  103.25 & $0.803\pm0.029$ \\
$\zeta$ Tau  & 53709.840  & 0.868 & W1/W2 &  104.15 &  100.64 & $0.720\pm0.019$ \\
$\kappa$ Dra & 53462.655  & 0.074 & S1/S2 &\phn26.40&\phn29.89& $0.929\pm0.037$ \\
$\kappa$ Dra & 53462.756  & 0.076 & S1/S2 &\phn27.74&\phn\phn4.61& $0.995\pm0.021$ \\
$\kappa$ Dra & 53462.774  & 0.076 & S1/S2 &\phn27.77&\phn\phn0.27& $0.997\pm0.022$ \\
$\kappa$ Dra & 53462.801  & 0.077 & S1/S2 &\phn27.70&  353.51 & $1.002\pm0.025$ \\
$\kappa$ Dra & 53462.827  & 0.077 & S1/S2 &\phn27.51&  346.82 & $1.058\pm0.041$ \\
$\kappa$ Dra & 53462.845  & 0.077 & S1/S2 &\phn27.29&  342.31 & $0.990\pm0.023$ \\
$\kappa$ Dra & 53462.859  & 0.078 & S1/S2 &\phn27.09&  338.86 & $1.019\pm0.034$ \\
$\kappa$ Dra & 53462.909  & 0.078 & S1/S2 &\phn26.04&  326.47 & $1.018\pm0.028$ \\
$\kappa$ Dra & 53462.926  & 0.079 & S1/S2 &\phn25.55&  322.13 & $0.974\pm0.035$ \\
$\kappa$ Dra & 53462.946  & 0.079 & S1/S2 &\phn24.92&  317.18 & $1.027\pm0.025$ \\
$\kappa$ Dra & 53462.963  & 0.079 & S1/S2 &\phn24.27&  312.66 & $0.979\pm0.021$ \\
$\kappa$ Dra & 53462.982  & 0.080 & S1/S2 &\phn23.50&  307.78 & $1.045\pm0.024$ \\
$\kappa$ Dra & 53462.999  & 0.080 & S1/S2 &\phn22.79&  303.56 & $0.999\pm0.042$ \\
$\kappa$ Dra & 53466.750  & 0.141 & S1/E1 &  267.57 &\phn43.26& $0.686\pm0.022$ \\
$\kappa$ Dra & 53466.847  & 0.142 & S1/E1 &  286.48 &\phn16.95& $0.738\pm0.032$ \\
$\kappa$ Dra & 53466.870  & 0.143 & S1/E1 &  288.46 &\phn10.67& $0.727\pm0.032$ \\
$\kappa$ Dra & 53466.890  & 0.143 & S1/E1 &  289.41 &\phn\phn5.37& $0.714\pm0.031$ \\
$\kappa$ Dra & 53466.914  & 0.144 & S1/E1 &  289.72 &  358.90 & $0.681\pm0.033$ \\
$\kappa$ Dra & 53473.724  & 0.254 & S1/E1 &  265.84 &\phn44.86& $0.776\pm0.029$ \\
$\kappa$ Dra & 53708.995  & 0.076 & W1/W2 &  104.62 &  154.45 & $0.832\pm0.027$ \\
$\kappa$ Dra & 53709.008  & 0.077 & W1/W2 &  104.95 &  149.84 & $0.855\pm0.057$ \\
$\kappa$ Dra & 53709.020  & 0.077 & W1/W2 &  105.29 &  145.59 & $0.847\pm0.055$ \\
$\kappa$ Dra & 53709.868  & 0.090 & W1/W2 &  104.32 &  200.84 & $0.817\pm0.026$ \\
$\kappa$ Dra & 53709.883  & 0.091 & W1/W2 &  104.05 &  195.44 & $0.713\pm0.018$ \\
$\kappa$ Dra & 53709.897  & 0.091 & W1/W2 &  103.85 &  190.28 & $0.753\pm0.023$ \\
$\kappa$ Dra & 53709.910  & 0.091 & W1/W2 &  103.73 &  185.24 & $0.859\pm0.021$ \\
$\kappa$ Dra & 53709.925  & 0.091 & W1/W2 &  103.69 &  179.64 & $0.835\pm0.018$ \\
$\kappa$ Dra & 53709.937  & 0.092 & W1/W2 &  103.73 &  175.03 & $0.760\pm0.017$ \\
$\kappa$ Dra & 53709.950  & 0.092 & W1/W2 &  103.83 &  170.33 & $0.719\pm0.015$ \\
$\kappa$ Dra & 53709.962  & 0.092 & W1/W2 &  104.00 &  165.71 & $0.799\pm0.018$ \\
$\kappa$ Dra & 53709.975  & 0.092 & W1/W2 &  104.21 &  161.15 & $0.873\pm0.015$ \\
$\kappa$ Dra & 53709.987  & 0.092 & W1/W2 &  104.48 &  156.65 & $0.889\pm0.018$ \\
$\kappa$ Dra & 53710.000  & 0.093 & W1/W2 &  104.80 &  151.94 & $0.926\pm0.022$ \\
$\kappa$ Dra & 53710.031  & 0.093 & W1/W2 &  105.70 &  140.57 & $0.879\pm0.026$ \\
$\kappa$ Dra & 53710.043  & 0.093 & W1/W2 &  106.07 &  136.19 & $0.871\pm0.029$ \\
\enddata
\tablenotetext{a}{The formal visibility errors listed here may underestimate the 
actual errors by a factor of $\approx 2.8$ (\S2.2).}
\end{deluxetable}

\clearpage


\begin{deluxetable}{lccccl}
\rotate
\tablewidth{0pc}
\tablecaption{Spectroscopic Observations\label{tab5}}
\tablehead{
\colhead{} &
\colhead{} &
\colhead{Resolving} &
\colhead{Range} &
\colhead{Date} &
\colhead{} \\
\colhead{Telescope} &
\colhead{Spectrograph} &
\colhead{Power} &
\colhead{(\AA )} &
\colhead{(UT)} &
\colhead{Targets}
}
\startdata
KPNO coud\'{e} feed 0.9 m   & Coude    &      12800 & 4235--4586 & 2004 Aug 16--21 
  & $\gamma$ Cas, $\phi$ Per \\
KPNO coud\'{e} feed 0.9 m   & Coude    &      12800 & 4236--4587 & 2005 Nov 10--14 
  & $\gamma$ Cas, $\phi$ Per, $\zeta$ Tau, $\kappa$ Dra \\
KPNO coud\'{e} feed 0.9 m   & Coude    &   \phn9150 & 6465--7140 & 2004 Oct 12--14 
  & $\gamma$ Cas, $\phi$ Per, $\zeta$ Tau \\
McDonald Obs. 2.7 m     & CoolSpec &   \phn1520 &20691--22520& 2004 Feb 06
  & $\gamma$ Cas, $\phi$ Per, $\zeta$ Tau \\
McDonald Obs. 2.7 m     & LCS      &\phn\phn810 & 3020--5717 & 2005 Apr 29
  & $\kappa$ Dra \\
Ond\v{r}ejov Obs. 2.0 m & Coude    &      12800 & 6262--6774 & 2005 Apr 11
  & $\kappa$ Dra \\
Ritter Obs. 1.0 m       & Echelle  &      26000 & 5413--6596 & 2005 Apr 10
  & $\kappa$ Dra \\
\enddata
\end{deluxetable}

\clearpage


\begin{deluxetable}{lcccccccc}
\tabletypesize{\scriptsize}
\rotate
\tablewidth{0pc}
\tablecaption{Thick Disk Model Fits of Visibilities\label{tab6}}
\tablehead{
\colhead{} &
\colhead{$\gamma$ Cas} &
\colhead{$\gamma$ Cas} &
\colhead{$\phi$ Per} &
\colhead{$\phi$ Per} &
\colhead{$\zeta$ Tau} &
\colhead{$\zeta$ Tau} &
\colhead{$\kappa$ Dra} &
\colhead{$\kappa$ Dra} \\
\colhead{Parameter} &
\colhead{Single} &
\colhead{Binary} &
\colhead{Single} &
\colhead{Binary} &
\colhead{Single} &
\colhead{Binary} &
\colhead{Single} &
\colhead{Binary}
}
\startdata
$i$ (deg)   \dotfill                   & 
 $51\pm4$      & $127\pm4$     & $90\pm7$      & $69\pm5$      &
 $90\pm3$      & $89\pm2$      & $72\pm18$     & $26\pm9$      \\        
$\alpha$ (deg)  \dotfill               &
 $116\pm4$     & $114\pm3$     & $46\pm4$      & $49\pm3$      &
 $38\pm2$      & $37\pm2$      & $114\pm4$     & $21\pm3$      \\        
$\log \rho_{0~{\rm fit}}$ (g cm$^{-3}$)  \dotfill  &
$-10.14\pm0.24$&$-10.18\pm0.25$&$-10.92\pm0.10$&$-10.98\pm0.08$&
 $-9.71\pm0.31$& $-9.73\pm0.35$&$-12.55\pm0.39$&$-12.21\pm0.32$\\        
$\log \rho_{0~{\rm IRAS}}$\tablenotemark{a} (g cm$^{-3}$)  \dotfill  &
$-10.95\pm0.35$&$-10.95\pm0.35$&$-10.85\pm0.25$&$-10.85\pm0.25$&
$-11.15\pm0.25$&$-11.15\pm0.25$&$-11.35\pm0.15$&$-11.35\pm0.15$\\        
$n_{\rm fit}$       \dotfill                     & 
 $2.70\pm0.31$ & $2.65\pm0.32$ & $1.80\pm0.09$ & $1.76\pm0.08$ &
 $3.14\pm0.38$ & $3.19\pm0.40$ & $0.16\pm0.43$ & $0.67\pm0.36$ \\        
$n_{\rm IRAS}$\tablenotemark{a}     \dotfill                       & 
  3.25         &  3.25         &  3.0          & 3.0           &
  3.25         &  3.25         &  3.0          & 3.0           \\        
$\chi^2_\nu$      \dotfill             &
 24.3          & 24.0          & 8.0           & 7.1           &
 7.9           & 7.9           & 4.0           & 4.0           \\        
$E(V-K)_{\rm fit}$ (mag)  \dotfill     &
 1.60          & 1.53          & 0.58          & 0.63          &
 0.53          & 0.40          & 0.31          & 0.14          \\        
$E(V-K)_{\rm obs}$\tablenotemark{b} (mag) \dotfill &
 0.85          & 0.85          & 0.68          & 0.68          &
 0.65          & 0.65          & 0.39          & 0.39          \\        
$\theta_{LD}$ (mas)    \dotfill        &
 0.50          & 0.50          & 0.30          & 0.30          &
 0.40          & 0.40          & 0.39          & 0.39          \\        
Disk FWHM (mas)   \dotfill  &
 1.35          & 1.36          & 0.98          & 0.51          &
 1.99          & 1.82          & 3.02          & 1.72          \\        
\enddata
\tablenotetext{a}{\citet{wat87}.}
\tablenotetext{b}{\citet{dou94}.}
\end{deluxetable}

\clearpage


\begin{deluxetable}{lcccc}
\tablewidth{0pc}
\tablecaption{Gaussian Elliptical Fits of Visibilities\label{tab7}}
\tablehead{
\colhead{Parameter} &
\colhead{$\gamma$ Cas} &
\colhead{$\phi$ Per} &
\colhead{$\zeta$ Tau} &
\colhead{$\kappa$ Dra} 
}
\startdata
$\theta_{\rm GD~fit}$ (mas) \dotfill &
 $1.95\pm0.07$ & $2.30\pm0.08$ & $1.79\pm0.07$ & $1.83\pm0.11$  \\
$\theta_{{\rm GD~H}\alpha}$\tablenotemark{a} (mas) \dotfill &
 $3.47\pm0.02$ & $2.67\pm0.20$ & $4.53\pm0.52$ & \nodata        \\
$\theta_{{\rm GD~H}\alpha}$\tablenotemark{b} (mas) \dotfill &
 $3.59\pm0.04$ & $2.89\pm0.09$ & $3.14\pm0.21$ & \nodata        \\
$r_{\rm fit}$  \dotfill    &
 $0.59\pm0.04$ & $0.00\pm0.22$ & $0.09\pm0.22$ & $0.00\pm0.67$  \\
$r_{{\rm H}\alpha}$\tablenotemark{a}  \dotfill    &
 $0.70\pm0.02$ & $0.46\pm0.04$ & $0.28\pm0.02$ & \nodata        \\
$r_{{\rm H}\alpha}$\tablenotemark{b}  \dotfill    &
 $0.58\pm0.03$ & $0.27\pm0.01$ & $0.31\pm0.07$ & \nodata        \\
$\phi_{\rm fit}$ (deg) \dotfill &
 $25\pm4$      & $-44\pm3$     & $-52.2\pm 1.7$& $-8\pm9$       \\
$\phi_{{\rm H}\alpha}$\tablenotemark{a} (deg) \dotfill &
 $19\pm2$      & $-62\pm5$     & $-58\pm4$     & \nodata        \\
$\phi_{{\rm H}\alpha}$\tablenotemark{b} (deg) \dotfill &
 $31.2\pm1.2$  & $-61.5\pm0.6$ & $-62.3\pm4.4$ & \nodata        \\
$c_{\rm P~fit}$ \dotfill  &
$0.395\pm0.018$&$0.538\pm0.015$&$0.414\pm0.029$&$0.628\pm0.030$ \\
$\chi^2_\nu$      \dotfill             &
 23.6          & 7.9           & 7.7           & 8.6            \\
\enddata
\tablenotetext{a}{\citet{qui97}.}
\tablenotetext{b}{\citet{tyc04,tyc06}.}
\end{deluxetable}

\clearpage


\begin{deluxetable}{lcccc}
\tablewidth{0pc}
\tablecaption{Emission Line Equivalent Widths and Density Fits\label{tab8}}
\tablehead{
\colhead{Parameter} &
\colhead{$\gamma$ Cas} &
\colhead{$\phi$ Per} &
\colhead{$\zeta$ Tau} &
\colhead{$\kappa$ Dra} 
}
\startdata
$W_\lambda ({\rm H}\alpha)$ (\AA )  \dotfill   &
 $-31.2$       & $-42.6$       & $-19.5$       & $-20.3$        \\
$W_\lambda ({\rm H}\gamma)$ (\AA )  \dotfill   &
\phn $+1.0$    & \phn $+2.1$   & \phn $+4.0$   & \phn $+5.9$    \\
$W_\lambda ({\rm Br}\gamma)$ (\AA )  \dotfill  &
\phn $-5.8$    & \phn $-9.2$   & \phn $-2.8$   & \nodata        \\
$\log \rho_{0~{\rm fit}} ({\rm H}\alpha)$ (g cm$^{-3}$)  \dotfill  &
$-11.7$        & $-12.5$       & $-11.2$       & $-13.4$        \\
$\log \rho_{0~{\rm fit}} ({\rm H}\gamma)$ (g cm$^{-3}$)  \dotfill  &
$-11.7$        & $-12.8$       & $-11.2$       & $-13.7$        \\ 
$\log \rho_{0~{\rm fit}} ({\rm Br}\gamma)$ (g cm$^{-3}$)  \dotfill  &
$-12.0$        & $-12.9$       & $-11.7$       & \nodata        \\
$\log \rho_{0~{\rm fit}} (K^\prime)$ (g cm$^{-3}$)  \dotfill  &
$-10.1\pm0.2$  & $-11.0\pm0.1$ & $-9.7\pm0.3$  & $-12.2\pm0.3$  \\        
\enddata
\end{deluxetable}




\clearpage

\clearpage

\clearpage

\clearpage

\clearpage

\clearpage

\clearpage

\clearpage

\clearpage

\clearpage

\clearpage

\clearpage

\clearpage

\clearpage

\begin{thebibliography}{}
\bibitem[Abt et al.(2002)Abt, Levato, \& Grosso]{abt02}
         Abt, H. A, Levato, H., \& Grosso, M. 2002, \apj, 573, 359 
\bibitem[Allende Prieto \& Lambert(1999)]{all99}     
	 Allende Prieto, C., \& Lambert, D. L. 1999, \aap, 352, 555
\bibitem[Aufdenberg et al.(2006)]{auf06}
         Aufdenberg, J. P., et al. 2006, \apj, 645, 664
\bibitem[Berger et al.(2006)]{ber06}
         Berger, D. H., et al. 2006, \apj, 644, 475
\bibitem[Berio et al.(1999)]{ber99}
         Berio, P., et al. 1999, \aap, 345, 203
\bibitem[Bjorkman \& Bjorkman(1994)]{bjo94}
         Bjorkman, J. E., \& Bjorkman, K. S. 1994, \apj, 436, 818
\bibitem[Carciofi \& Bjorkman(2006)]{car06}
         Carciofi, A. C., \& Bjorkman, J. E.
         2006, \apj, 639, 1081
\bibitem[Cassinelli et al.(1987)Cassinelli, Nordsieck, \& Murison]{cas87}
         Cassinelli, J. P., Nordsieck, K. H., \& Murison, M. A.
         1987, \apj, 317, 290
\bibitem[Cayrel de Strobel et al.(1997)]{cay97}
         Cayrel de Strobel, G., Soubiran, C., Friel, E. D., Ralite, N.,
         \& Francois, P. 1997, \aap, 124, 299
\bibitem[Chesneau et al.(2005)]{che05} 
         Chesneau, O., et al. 2005, \aap, 435, 275
\bibitem[Claret(2000)]{cla00}
         Claret, A. 2000, \aap, 363, 1081
\bibitem[Clarke(1990)]{cla90}
         Clarke, D. 1990, \aap, 227, 151
\bibitem[Clarke \& Bjorkman(1998)]{cla98}
         Clarke, D., \& Bjorkman, K. S. 1998, \aap, 331, 1059
\bibitem[Cohen et al.(2003)Cohen, Wheaton, \& Megeath]{coh03}
         Cohen, M., Wheaton, W. A., \& Megeath, S. T.
         2003, \aj, 126, 1090
\bibitem[Colina et al.(1996)Colina, Bohlin, \& Castelli]{col96}
         Colina, L., Bohlin, R., \& Castelli, F. 1996, HST Instrument
         Science Report CAL/SCS-008 (Baltimore: STScI)
\bibitem[Cot\'{e} \& Waters(1987)]{cot87}
	 Cot\'{e}, J., \& Waters, L. B. F. M. 1987, \aap, 176, 93
\bibitem[Cutri et al.(2003)]{cut03}
         Cutri, R. M., et al. 2003,
         The 2MASS All-Sky Catalog of Point Sources
         (Pasadena: IPAC/Cal. Inst. Tech.)
\bibitem[Davis et al.(2000)Davis, Tango, \& Booth]{dav00}
         Davis, J., Tango, W. J., \& Booth, A. J.
         2000, \mnras, 318, 387
\bibitem[Domiciano de Souza et al.(2003)]{dom03}
         Domiciano de Souza, A., Kervella, P., Jankov, S., Abe, L.,
         Vakili, F., di Folco, E., \& Paresce, F. 
         2003, \aap, 407, L47
\bibitem[Doazan(1982)]{doa82}
         Doazan, V. 1982, in B Stars With and Without Emission Lines
         (NASA SP-456), ed A. Underhill \& V. Doazan (Washington, DC: NASA), 279
\bibitem[Dougherty \& Taylor(1992)]{dou92}
         Dougherty, S. M., \& Taylor, A. R.
         1992, Nature, 359, 808
\bibitem[Dougherty et al.(1994)]{dou94}
	 Dougherty, S. M., Waters, L. B. F. M., Burki, G., Cot\'{e}, J.,
         Cramer, N., van Kerkwijk, M. H., \& Taylor, A. R.
         1994, \aap, 290, 609
\bibitem[Fitzpatrick(1999)]{fit99}
         Fitzpatrick, E. L. 1999, \pasp, 111, 63
\bibitem[Gies(2000)]{gie00}
         Gies, D. R. 2000, in The Be Phenomenon in Early-Type Stars,
         IAU Coll. 175 (ASP Conf. Ser. 214), ed. M. Smith, H. Henrichs, 
         \& J. Fabregat (San Francisco: ASP), 668
\bibitem[Gies et al.(1998)]{gie98}
         Gies, D. R.. Bagnuolo, W. G., Jr., Ferrara, E. C., Kaye, A. B.,
         Thaller, M. L., Penny, L. R., \& Peters, G. J. 1998, \apj, 493, 440
\bibitem[Gies et al.(2002)]{gie02}
         Gies, D. R., McSwain, M. V., Riddle, R. L., Wang, Z., Wiita, P. J.,
         \& Wingert, D. W. 2002, \apj, 566, 1069
\bibitem[Gontcharov et al.(2000)Gontcharov, Andronova, \& Titov]{gon00}
         Gontcharov, G. A., Andronova, A. A., \& Titov, O. A.
         2000, \aap, 355, 1164
\bibitem[Gray(1998)]{gra98}
         Gray, R. O. 1998, \aj, 116, 482
\bibitem[Guo et al.(1995)]{guo95}
         Guo, Y., Huang, L., Hao, J., Cao, H., Guo, Z., \& Guo, X.
         1995, \aap, 112, 201
\bibitem[Hanuschik(1996)]{han96}
         Hanuschik, R. W. 1996, \aap, 308, 170
\bibitem[Harmanec(1984)]{har84}
         Harmanec, P. 1984, Bull. Astron. Inst. Czechoslovakia, 35, 164
\bibitem[Harmanec et al.(2000)]{har00}
         Harmanec, P., et al. 2000, \aap, 364, L85 
\bibitem[Hony et al.(2000)]{hon00}
         Hony, S., et al. 2000, \aap, 355, 187
\bibitem[Howells et al.(2001)]{how01}
	 Howells, L., Steele, I. A., Porter, J. M., \& Etherton, J.	
         2001, \aap, 369, 99
\bibitem[Hubeny \& Lanz(1995)]{hub95}
         Hubeny, I., \& Lanz, T. 1995, \apj, 439, 875
\bibitem[Hummel \& Dachs(1992)]{hum92}
         Hummel, W., \& Dachs, J. 1992, \aap, 262, L17
\bibitem[Hummel \& \v{S}tefl(2001)]{hum01}
         Hummel, W., \& \v{S}tefl, S. 2001, \aap, 368, 471
\bibitem[Hummel \& Vrancken(1995)]{hum95}
         Hummel, W., \& Vrancken, M. 1995, \aap, 302, 751
\bibitem[Hummel \& Vrancken(2000)]{hum00}
         Hummel, W., \& Vrancken, M. 2000, \aap, 359, 1075
\bibitem[Juza et al.(1991)]{juz91}
         Juza, K., Harmanec, P., Hill, G. M., Tarasov, A. E., \& Matthews, J. M.
         1991, Bull. Astron. Inst. Czechoslovakia, 42, 39
\bibitem[Kaye \& Gies(1997)]{kay97}
         Kaye, A. B., \& Gies, D. R. 1997, \apj, 482, 1028
\bibitem[Kharitonov et al.(1988)Kharitonov, Tereshchenko, \& Knyazeva]{kha88}
         Kharitonov, A. V., Tereshchenko, V. M., \& Knyazeva, L. N.
         1988, Alma-Ata: Nauka, 484
\bibitem[Kraus et al.(2005)]{kra05}
         Kraus, S., et al. 2005, \aj, 130, 246
\bibitem[Lambert \& Reddy(2004)]{lam04}
         Lambert, D. L., \& Reddy, B. E. 2004, \mnras, 349, 757
\bibitem[Lester et al.(2000)]{les00}
         Lester, D. F., Hill, G. J., Doppmann, G., \& Froning, C. S.
         2000, \pasp, 112, 384
\bibitem[Likkel et al.(2006)]{lik06}
         Likkel, L., Dinerstein, H. L., Lester, D. F., Kindt, A., \& Bartig, K.
         2006, \aj, 131, 1515
\bibitem[Mason et al.(1997)]{mas97}
         Mason, B. D., ten Brummelaar, T., Gies, D. R., Hartkopf, W. I.,
         \& Thaller, M. L. 1997, \aj, 114, 2112
\bibitem[McAlister et al.(2005)]{mca05}
         McAlister, H. A., et al. 2005, \apj, 628, 439
\bibitem[McSwain \& Gies(2005)]{mcs05}
	 McSwain, M. V., \& Gies, D. R. 2005, \apjs, 161, 118
\bibitem[Miroshnichenko et al.(2002)Miroshnichenko, Bjorkman, \& Krugov]{mir02}
         Miroshnichenko, A. S., Bjorkman, K. S., \& Krugov, V. D. 
         2002, PASP, 114, 1226
\bibitem[Mourard et al.(1989)]{mou89}
	 Mourard, D., Bosc, I., Labeyrie, A., Koechlin, L., \& Saha, S. 
         1989, Nature, 342, 520
\bibitem[Nordstr\"{o}m et al.(2004)]{nor04}
         Nordstr\"{o}m, B., et al. 2004, \aap, 418, 989
\bibitem[Okazaki et al.(2002)]{oka02}
         Okazaki, A. T., Bate, M. R., Ogilvie, G. I., \& Pringle, J. E.
         2002. \mnras, 337, 967
\bibitem[Owocki(2005)]{owo05}
         Owocki, S. 2005, in The Nature and Evolution of Disks 
         Around Hot Stars (ASP Conf. Ser. 337), 
         ed. R. Ignace \& K. G. Gayley (San Francisco: ASP), 101
\bibitem[Perryman(1997)]{per97}
         Perryman, M. A. C. 1997, The Hipparcos and Tycho
         Catalogues, ESA SP-1200 (Noordwijk: ESA/ESTEC)
\bibitem[Poeckert \& Marlborough(1979)]{poe79}
         Poeckert, R., \& Marlborough, J. M. 1979, \apj, 233, 259
\bibitem[Pols et al.(1991)]{pol91}
         Pols, O. R., Cot\'{e}, J., Waters, L. B. F. M., \& Heise, J. 
         1991, \aap, 241, 419
\bibitem[Porter \&  Rivinius(2003)]{por03}
         Porter, J. M., \&  Rivinius, Th. 2003, \pasp, 115, 1153
\bibitem[Quirrenbach et al.(1997)]{qui97}
         Quirrenbach, A., et al. 1997, \apj, 479, 477
\bibitem[Rinehart et al.(1999)Rinehart, Houck, \& Smith]{rin99}
         Rinehart, S. A., Houck, J. R., \& Smith, J. D. 1999, \aj, 118, 2974
\bibitem[Robinson et al.(2002)Robinson, Smith, \& Henry]{rob02}
         Robinson, R. D., Smith, M. A., \& Henry, G. W.
         2002, \apj, 575, 435
\bibitem[Saad et al.(2004)]{saa04}
         Saad, S. M., et al. 2004, \aap, 419, 607
\bibitem[Saad et al.(2005)]{saa05}
         Saad, S. M., et al. 2005, \apss, 296, 173
\bibitem[Smith et al.(2004)Smith, Price, \& Baker]{smi04}
         Smith, B. J., Price, S. D., \& Baker, R. I. 2004, 
         \apjs, 154, 673 
\bibitem[Stee(2003)]{ste03}
         Stee, Ph. 2003, \aap, 403, 1023
\bibitem[Stee \& Bittar(2001)]{ste01}
         Stee, Ph., \& Bittar, J. 2001, \aap, 367, 532
\bibitem[Stee et al.(1995)]{ste95}
         Stee, Ph., de Araujo, F. X., Vakili, F., Mourard, D., 
         Arnold, L., Bonneau, D., Morand, F., \& Tallon-Bosc, I.
         1995, \aap, 300, 219 
\bibitem[Stee et al.(2005)]{ste05}
         Stee, Ph., Meilland, A., Berger, D., \& Gies, D.
         2005, in The Nature and Evolution of Disks Around Hot Stars
         (ASP Conf. Ser. 337), ed. R. Ignace \& K. G. Gayley
         (San Francisco: ASP), 211
\bibitem[Sturmann et al.(2003)]{stu03}
         Sturmann, J., ten Brummelaar, T. A., Ridgway, S. T., 
         Shure, M. A., Safizadeh, N., Sturmann, L., Turner, N. H., 
         \& McAlister, H. A. 2003, Proc. SPIE, 4838, 1208
\bibitem[ten Brummelaar et al.(2005)]{ten05}
         ten Brummelaar, T. A., et al. 2005, \apj, 628, 453
\bibitem[Thom et al.(1986)Thom, Granes, \& Vakili]{tho86}
         Thom, C., Granes, P., \& Vakili, F. 1986, \aap, 165, L13
\bibitem[Townsend et al.(2004)Townsend, Owocki, \& Howarth]{tow04}
	 Townsend, R. H. D., Owocki, S. P., \& Howarth, I. D.
         2004, \mnras, 350, 189
\bibitem[Tycner et al.(2003)]{tyc03}
         Tycner, C., et al. 2003, \aj, 125, 3378
\bibitem[Tycner et al.(2004)]{tyc04}
         Tycner, C., et al. 2004, \aj, 127, 1194
\bibitem[Tycner et al.(2005)]{tyc05}
         Tycner, C., et al. 2005, \apj, 624, 359
\bibitem[Tycner et al.(2006)]{tyc06}
         Tycner, C., et al. 2006, \aj, 131, 2710
\bibitem[Vakili et al.(1998)]{vak98}
         Vakili, F., et al. 1998, \aap, 335, 261
\bibitem[van Belle \& van Belle(2005)]{van05}
         van Belle, G. T., \& van Belle, G. 2005, \pasp, 117, 1263
\bibitem[Wallace \& Hinkle(1997)]{wal97}
         Wallace, L., \& Hinkle, K. 1997, \apjs, 111, 445
\bibitem[Waters(1986)]{wat86}
	 Waters, L. B. F. M. 1986, \aap, 162, 121
\bibitem[Waters \& Lamers(1984)]{wat84}
	 Waters, L. B. F. M., \& Lamers, H. J. G. L. M. 1984, \aaps, 57, 327
\bibitem[Waters et al.(1987)Waters, Cot\'{e}, \& Lamers]{wat87}
	 Waters, L. B. F. M., Cot\'{e}, J., \& Lamers, H. J. G. L. M.
         1987, \aap, 185, 206
\bibitem[Yang et al.(1990)]{yan90}
         Yang, S., Walker, G. A. H., Hill, G. M., \& Harmanec, P.
         1990, \apjs, 74, 595
\bibitem[Yoo et al.(2002)Yoo, Bak, \& Lee]{yoo02}
         Yoo, J. J., Bak, J.-Y., \& Lee, H.-W. 2002, \mnras, 336, 467
\bibitem[Zorec et al.(2005)Zorec, Fr\'{e}mat, \& Cidale]{zor05}
	 Zorec, J., Fr\'{e}mat, Y., \& Cidale, L. 2005, \aap, 441, 235
\end{thebibliography}
\end{document}